\documentclass[12pt,a4paper]{article}
\usepackage[UKenglish]{babel}
\usepackage[utf8]{inputenc}
\usepackage[T1]{fontenc}
\usepackage[top=2.5cm, bottom=2.5cm, left=2.0cm, right=2.0cm]{geometry}

\usepackage{amscd,amsfonts,amsmath,amssymb,amsthm,bbm,bm,latexsym,mathrsfs,mathtools,thmtools}
\usepackage[subnum]{cases}	% multi-case equations with separate number for each
\makeatletter
\newcommand*{\distas}[1]{\mathbin{\overset{#1}{\kern\z@\sim}}}	
\makeatother
\newcommand*{\norm}[1]{\left\lVert#1\right\rVert}		% norm
\newcommand*\abs[1]{\left|#1\right|}		% absolute value
\allowdisplaybreaks			% breaks "align" in multiple pages

\newtheoremstyle{custom}%    <name>
                {\topsep}%   <space above>
                {\topsep}%   <space below>
                {\itshape}%  <body font>
                {}%          <indent amount>
                {\bfseries}% <Theorem head font>
                {.}%         <punctuation after theorem head>
                {\newline}%  <space after theorem head> (default .5em)
                {}%          <Theorem head spec>
\theoremstyle{custom}
\newtheorem{theorem}{Theorem}[section]

\newtheorem{proposition}{Proposition}[section]

\theoremstyle{remark}
\newtheorem{remark}{Remark}

\theoremstyle{plain}

\newtheorem{example}{Example}

\usepackage{hyperref} 		% pagebackref=true, linktoc=all, colorlinks=true, citecolor=red
\hypersetup{bookmarks=true, unicode=true, urlcolor=blue, linkcolor=blue, filecolor=blue, colorlinks=true, citecolor=black, final=true}

\usepackage{xcolor,subfigure,epsfig,graphics,graphicx} \graphicspath{{./images/}}
\usepackage{caption}
\usepackage{tikz,tkz-berge}
\usetikzlibrary{shapes,arrows,bayesnet}	% flow charts
\definecolor{darkgreen}{rgb}{0.0,0.5,0.0}  %{0.0,0.26,0.15}
\definecolor{dgreen}{RGB}{0,100,0}
\definecolor{darkred}{rgb}{0.8,0,0}
\colorlet{dred}{darkred!80!black}
\definecolor{darkblue}{RGB}{25,25,160}
\definecolor{orange}{RGB}{255,90,0}
\usepackage{natbib}
\usepackage{enumitem}

\usepackage{longtable,float,array}
\usepackage{setspace}

\makeatother
\usepackage{lineno}					% numbered lines in text

\title{\vspace{-50pt} \textbf{Bayesian nonparametric graphical models for time-varying parameters VAR}\thanks{We are grateful to Deborah Gefang, Roberto Léon-Gonzalez and Sylvia Kaufmann for their comments and suggestions. Moreover, we thank the participants at: ``4th Vienna Workshop on High-dimensional Time Series in Macroeconomics and Finance'' in Wien, 2019. This research used the SCSCF multiprocessor cluster system at Ca' Foscari University of Venice. \newline
Luca Rossini acknowledges financial support from the European Union Horizon 2020 research and innovation programme under the Marie Sklodowska-Curie grant agreement No 796902.}}

\author{
\hspace{-20pt}
Matteo Iacopini\textsuperscript{a}\thanks{e-mail: \href{mailto: matteo.iacopini@sns.it}{matteo.iacopini@sns.it}} \hspace{20pt} Luca Rossini\textsuperscript{b}\thanks{e-mail: \href{mailto: l.rossini@vu.nl}{l.rossini@vu.nl}}
 \\
 \\
{\centering {\small
\textsuperscript{a}Scuola Normale Superiore of Pisa, Italy \hspace{10pt} \textsuperscript{b}Vrije Universiteit Amsterdam, The Netherlands}}
}

\date{\normalsize \today}

\begin{document}

\maketitle
\singlespacing

%\abstract{
\begin{abstract}
Over the last decade, big data have poured into econometrics, demanding new statistical methods for analysing high-dimensional data and complex non-linear relationships. A common approach for addressing dimensionality issues relies on the use of static graphical structures for extracting the most significant dependence interrelationships between the variables of interest. Recently, Bayesian nonparametric techniques have become popular for modelling complex phenomena in a flexible and efficient manner, but only few attempts have been made in econometrics.

In this paper, we provide an innovative Bayesian nonparametric (BNP) time-varying graphical framework for making inference in high-dimensional time series. We include a Bayesian nonparametric dependent prior specification on the matrix of coefficients and the covariance matrix by mean of a Time-Series DPP as in \cite{NietoBarajas12TimeSeries_DDP}. Following \cite{Billio19BNP_sparseVAR}, our hierarchical prior overcomes over-parametrization and over-fitting issues by clustering the vector autoregressive  (VAR) coefficients into groups and by shrinking the coefficients of each group toward a common location. %Thus this hierarchical prior allows to contemporaneously estimate the (potentially) sparse time-varying causal network structure and to cluster the corresponding coefficients.
Our BNP time-varying VAR model is based on a spike-and-slab construction coupled with dependent Dirichlet Process prior (DPP) and allows to: (i) infer time-varying Granger causality networks from time series; (ii) flexibly model and cluster non-zero time-varying coefficients; (iii) accommodate for potential non-linearities.

In order to assess the performance of the model, we study the merits of our approach by considering a well-known macroeconomic dataset. Moreover, we check the robustness of the method by comparing two alternative specifications, with Dirac and diffuse spike prior distributions.

\vspace*{10pt}
\noindent \textbf{Keywords:} Bayesian Nonparametrics; Dependent Dirichlet process; Large vector autoregression; Sparsity; Time-Varying networks.\\[2pt]
\textbf{AMS 2000 subject classifications:} Primary 62; secondary 91B84.\\[2pt]
\textbf{JEL Classification:} C11, C32, C51, C53
\end{abstract}
%}

\newpage
\singlespacing

\section{Introduction}
Over the last decade, the availability of large datasets in economics and finance has allowed the introduction of high dimensional models. In particular, large datasets in macroeconomics help to improve the forecasts, while in finance some authors have investigated the use of large datasets to analyse financial crises, contagion effects and their impact on the real economy. In order to deal with high dimensional models, the introduction of Bayesian nonparametric techniques have become popular in different fields (such as statistics and machine learning), but only few attempts have been made in econometrics. In particular, Bayesian nonparametric approach allows to improve the estimation efficiency and the prediction accuracy in time series analysis.

Recently, time-varying parameter (TVP) models provide an interesting alternative to process multiple change points; for example, time-varying structural vector autoregressive (VAR) models have been used in \cite{Primiceri05Bayes_TVP_VAR} for study monetary policy application; \cite{Dangl12predictive_TVP} forecast equity returns by mean of TVP models; and in \cite{Belmonte14shrink_TVP} the European inflation has been studied via a time-varying parameters model. As shown in \cite{Primiceri05Bayes_TVP_VAR},  \citet{DelNegroPrimiceri15Bayes_TVP_VAR} and \cite{Bitto19Shrinkage_Bayes_TVP_VAR}, the advantage in capturing gradual changes is due to the flexibility of TVP models. We combine the ideas behind time-varying parameters models and Bayesian nonparametric techniques, thus allowing to model  complex phenomena in a flexible and efficient manner. Moreover, we provide an innovative Bayesian nonparametric time-varying graphical framework for making inference in high-dimensional time series. 

In this paper, we allow coefficients to be sparse, meaning that only a fraction of the time varying parameters have significant effects, but we retain flexibility in modelling non-zero coefficients, by including temporal dependence in the prior structure. In order to achieve these goals, we define a shrinkage prior on the VAR coefficients by means of a Bayesian nonparametric prior (BNP). This distribution is a spike-and-slab prior, where on the spike (parametric) distribution, we impose two different specifications: a Dirac spike and a ``diffuse'' spike. On the other hand, on the (non-parametric) slab distribution, we use a well know Bayesian nonparametric Lasso prior as in \cite{Billio19BNP_sparseVAR}. 

The prior previously described groups the time-varying parameter vector autoregressive (TVP-VAR) coefficients into clusters and shrinks the coefficients within a cluster toward common notation. Differently from Markov-switching approach \citep{Krolzig97MS} and random walk processes \citep{Primiceri05Bayes_TVP_VAR,DelNegroPrimiceri15Bayes_TVP_VAR}, we impose time variation on the distribution of the VAR coefficients. In the literature of time-varying coefficients, the VAR coefficients are represented as a direct dependence, in practice they can be represented as state-space models, where they are functions of the previous time. On the other hand, we introduce a different structure, the indirect dependence on the VAR coefficients. In this case, we have a dependence construct through the atoms of the Dirichlet process and not on a direct way.
%In our approach, we have an indirect dependence on the VAR coefficients though the atoms of the Dirichlet process and not a direct dependence based on a state-space models. 
Thus, we include a Bayesian nonparametric dependent prior specification on the VAR coefficients and the covariance matrix by means of a time-series dependent Dirichlet process (tsDDP) as in \cite{NietoBarajas12TimeSeries_DDP}. 

Following \cite{Billio19BNP_sparseVAR}, our hierarchical prior overcomes overparametrization and overfitting issues by clustering the VAR coefficients into groups and by shrinking the coefficients of each group toward a common location. This hierarchical prior allows to contemporaneously estimate the (potentially) sparse time-varying causal network structure and to cluster the corresponding coefficients.
In our BNP-TVP-VAR model, time-varying coefficients allow to (i) estimate the temporal networks of contemporaneous and causal structures, (ii) identify different sources of time variation, from the size of shocks and/or the propagation mechanism, and (iii) accommodate for potential non-linearities.

We also contribute to the literature on financial and macroeconomic contagion (see \cite{Billioetal12GrangerNet}; \cite{Bianchi19GraphicalSUR} and \cite{Barigozzi19NETS_network_estimation}) through the lens of Granger causality and graphs/networks representation. Our BNP prior is particularly suited for studying Granger causality from time series and in particular it allows to estimate the most significant time-varying dependence interrelationships between the variables of interest. As explained above, we can extract time-varying graphs by using the posterior random partition induced by the non-parametric (slab) distribution, which allows to cluster the edges into groups.

%A common approach for addressing dimensionality issues relies on the use of static graphical structures for extracting the most significant dependence interrelationships between the variables of interest. 

%To assess how our model performs in a typical real-data application, we study the merits of our approach by considering the macroeconomic dataset of \cite{McCracken16FRED-MD_dataset}. 
%Following \cite{Giannone14priorVAR}, we use three sets of variables to estimate a small-scale, medium-scale, and a large-scale VAR. The small-scale VAR only includes the real GDP growth, the GDP deflator, and the Federal Funds Rate (FFR). The medium-scale VAR additionally covers data on investment and consumption growth and hours worked (a proxy for capacity utilization of the economy). These are typical variables that are used in estimating DSGE models in the spirit of \cite{Smets07BayesDSGE}. Finally, the large-scale VAR features 21 variables.

%We illustrate the benefits of our approach by extracting temporal network structures from economic panel data, for analysing and forecasting shock transmission in business cycles and in financial markets.

\subsection{Literature}
%The use of Bayesian nonparametrics in econometrics has experienced a wide increase over the last few years. 
Since their introduction in macroeconomics (see \cite{Sims80VAR_macroeconomics}), vector autoregessive (VAR) models have been extensively used in econometrics and time series statistics.
Large VAR models have been used to analyse and forecast high-dimensional macroeconomic data (e.g., \cite{McCracken16FRED-MD_dataset}) and financial panels (e.g., \cite{Barigozzi19NETS_network_estimation}). Moreover, in recent years VAR models have been used for studying financial and macroeconomic contagion (e.g., \cite{Cogley05Bayes_TVP_VAR_macro}, \cite{Stock07US_inflation_forecast}, \cite{Diebold12GVAR_VolatilitySpillover} and \cite{Bianchi19GraphicalSUR}). Although, VAR models have been extensively used for assessing the impact and spread of external shocks (i.e., to perform impulse-response analysis), forecasting, estimating networks from Granger-causal relationships and to study systemic risk and financial contagion (e.g., \cite{Diebold09VolatilitySpillover},  \cite{Billioetal12GrangerNet} and \cite{Barigozzi19NETS_network_estimation}).

Despite being a potentially very flexible statistical tools, the high number of parameters and the typical limited length of standard macroeconomic datasets make unrestricted inference daunting as the cross-sectional size increases. This has favoured the use of penalised regression and Bayesian methods for dealing with the problem of over-parametrisation. The general idea is to use informative priors to shrink the unrestricted model towards a more parsimonious setting, thereby reducing parameter uncertainty and improving forecast accuracy (see \cite{Karlsson13Forecasting_BVAR_survey}, \cite{Koop10BVAR_survey} for a survey).

In the Bayesian VAR (a.k.a. BVAR) literature, a plethora of different prior distributions have been proposed to perform sparse estimation (e.g., see \cite{Giannone14priorVAR}). Starting from the well-known Minnesota prior (see \cite{DoanLitterman84BVAR_MinnesotaPrior_survey}, \cite{Litterman86BVAR_MinnesotaPrior}), which specifies an objective prior on the coefficient and covariance matrices of a VAR, several parametric approaches have been developed exploiting hierarchical structures and finite mixtures (e.g., \cite{Kalli14Bayes_sparse_TVP_VAR}, \cite{Gefang14Bayes_DoubleElasticNet},  \cite{Huber19AdaptiveShrinkage_VAR}, \cite{Kastner18Sparse_largeVAR}). 

Among the recent contributions for dealing with large dimensional models, we distinguish two approaches: the first attempts to reduce the size of the data to handle or to process during each step of the inferential algorithm, while the second is concerned with the reduction of the size of the parameter space.
Within the first class, we mention the Bayesian compressed VAR of \cite{KoopKorobilis18BayesianCompressedVAR}, who tackled the dimensionality issue by using random projections to compress the data, and the Bayesian composite likelihood approach of \cite{Chan18CompLike_BayesVAR}.
On the other hand, \cite{Gefang19VariationalBayes_largeVAR} and \cite{Koop18VariationalBayes_VAR} adopted a variational Bayes approach for performing efficient approximate posterior inference in large parameter spaces. Also, \cite{Kastner18Sparse_largeVAR} exploited factor models and hierarchical shrinkage priors for providing a parsimonious parametrisation of the covariance matrix which allows for equation-by-equation estimation.
Additional contributions for estimating large VAR and VARMA models include \cite{Koop13Large_TVP_VAR}, \cite{Korobilis16PanelVAR_priorShrinkage} and \cite{Chan16Large_BayesVARMA}.

In addition to the large cross-sectional dimensionality, also the temporal length of many economic and financial datasets is steadily increasing. 
Thus, the possible relations between different variables of interest can be described by static matrix of coefficients. This assumption can be elapsed by introducing a time-variation of the matrix of coefficients of the time series. In particular, 
%This naturally leads to question whether the relation specified by the coefficients of the model is actually static or it varies through time. To answer this question, 
the most common approach consists in specifying a process governing the evolution of the parameters of interest. According to the force driving this dynamics, we distinguish observation-driven and parameter-driven time-varying parameter (TVP) models. The first class is mainly represented by generalised autoregressive score models (GAS, see \cite{Creal13GAS}), while the second one includes Markov switching (e.g., \cite{Hamilton89MS}, \cite{Krolzig97MS}), change point (e.g. \cite{Pesaran06ChangePoint_VAR}) and random walk models (e.g. \cite{DelNegroPrimiceri15Bayes_TVP_VAR}, \cite{Primiceri05Bayes_TVP_VAR}).
These processes are able to describe parameters whose evolution is subject to switching regimes, structural breaks or smooth changes, respectively.

In the Bayesian and frequentist literature, the use of parametric models has been widely studied by applying different shrinkage methods (such as the Least Absolute Shrinkage and Selection Operator, known as LASSO). In particular, important papers focus on  
%The vast majority of statistical tools used in the frequentist and Bayesian VAR literature are inherently parametric. This reflects the main focus on
sparse and efficient estimation in high-dimensional datasets. However, more recently increasing attention is being devoted to the issue of over-shrinkage and to the modelling of non-zero coefficients (e.g., \cite{Giannone18IllusionSparsity}).
Consequently, there is an increasing need for adequate statistical tools capable of flexibly model the dynamics described by a VAR process, allowing for sparsity without incurring into over-shrinking.

In this paper, we aim to contribute to the growing literature on the use of Bayesian nonparametrics in time series analysis. In particular, Bayesian nonparametric techniques are widely in statistics, machine learning and data analysis as powerful tools for flexible modelling of complex data structure. Only recently, Bayesian nonparametrics has increased popularity in econometrics and in economic time series modelling to capture
observation clustering effects (see e.g.  \cite{Bassetti14BetaPitmanYor},  \cite{Kalli18BNP_VAR} and \cite{Billio19BNP_sparseVAR}). 

%The last decade has experienced a remarkable development of Bayesian nonparametric techniques, which are now widely used in statistics, machine learning and data analysis as powerful tools for flexible modelling of complex data structures. Despite this increasing popularity, the exploitation of these techniques in econometrics is still at its infancy (e.g., see \cite{Billio19BNP_sparseVAR}, \cite{Kalli18BNP_VAR}, \cite{Bassetti14BetaPitmanYor}).

Up to our knowledge, our paper is the first to provide sparse Bayesian nonparametric VAR model when the coefficients are time-varying and the proposed two-stage prior specification can be easily extended to other classes, such as the seemingly unrelated regression (SUR) models. We propose a novel Bayesian nonparametric prior structure, which provides a sparse estimation of the coefficient matrix of a VAR model. This representation allows to manage the flexibility of non-zero entries and most importantly, to manage the time-variation in the matrix of coefficients through the atoms of the Dirichlet process and not through a state-space representation.

Our approach substantially differs from the existing literature in two aspects as described below. First, we consider a spike-and-slab prior distribution for each entry of the coefficient matrix, where on the spike we have a parametric prior specification by mean of Dirac or diffuse prior. On the hand,  the slab component has random nonparametric prior. Second, we impose prior dependence on the coefficients by specifying a Markov process for their random distribution. As a by-product of the estimation procedure, we are able to extract a time series of dependent Granger-causality graphs. This shows how the BNP-TVP-VAR contributes to the literature on the estimation of time-varying networks from economic and/or financial series.

The paper is organized as follows. Section \ref{sec:model} introduces the modelling framework and presents the BNP-TVP prior structure, then Section \ref{sec:posterior} presents posterior approximation and describes how to extract Granger-causal time varying graphs from time series. %We apply the methodology to study the FRED macroeconomic dataset in \autoref{sec:application}. 
Finally, Section \ref{sec:conclusions} draws the conclusions.

\section{A Bayesian Time-Varying VAR Model} \label{sec:model}

\subsection{TVP-VAR models}

Let $n$ be the number of units in a dataset and $\mathbf{y}_t = (y_{1,t},\dots,y_{n,t})$ a vector of $n$ variables available at time $t$. A time-varying parameters vector autoregressive model of order $p$ (TVP-VAR($p$)) is defined as
\begin{equation}
\mathbf{y}_t = \sum_{i=1}^p \mathbf{B}_{t-i+1} \mathbf{y}_{t-i} + \boldsymbol{\epsilon}_t, \qquad \boldsymbol{\epsilon}_t \sim \mathcal{N}(\mathbf{0},\boldsymbol{\Sigma}).
\label{eq:model_general}
\end{equation}
where $B_{t}$ is the $(n\times n)$ matrix of time-varying coefficients an $t=p,\ldots,T$ is the time period. We assume that the error terms $\boldsymbol{\epsilon}_t = (\epsilon_{1,t},\ldots, \epsilon_{n,t})'$ are i.i.d. for $t$ with Gaussian distribution $ \mathcal{N}(\mathbf{0},\boldsymbol{\Sigma})$.
Eq. \eqref{eq:model_general} can be written in the more compact form as 
\begin{equation}
\mathbf{y}_t = \sum_{i=1}^p \mathbf{X}_{t-i+1} \boldsymbol{\beta}_{t-i+1} + \boldsymbol{\epsilon}_t, \qquad \boldsymbol{\epsilon}_t \sim \mathcal{N}(\mathbf{0},\boldsymbol{\Sigma}).
\label{eq:model_general_vectorised}
\end{equation}
where we define $\boldsymbol{\beta}_t = \operatorname{vec}(\mathbf{B}_t)$; $\mathbf{X}_t = (\mathbf{y}_{t-1}' \otimes \mathbf{I}_n)$; $\otimes$ is the Kronecker product and $\operatorname{vec}(\cdot)$ the column-wise vectorization operator that stacks the columns of a matrix into a column vector.

%In the literature on macroeconomics and finance, VAR models have been extensively used for assessing the impact and spread of external shocks (i.e., to perform impulse-response analysis), forecasting, estimating networks from Granger-causal relationships and to study systemic risk and financial contagion (e.g., \cite{Diebold09VolatilitySpillover}, \cite{Barigozzi19NETS_network_estimation}).

\subsection{Prior specification}
Let us consider the problem of defining a flexible prior for a time-varying parameter model and we define the following TVP-VAR($p$) with $p$ equal to $1$ as
\begin{equation}
\mathbf{y}_t = \mathbf{X}_t \boldsymbol{\beta}_t + \boldsymbol{\epsilon}_t, \qquad \boldsymbol{\epsilon}_t \sim \mathcal{N}(\mathbf{0},\boldsymbol{\Sigma}).
\label{eq:model_vectorised}
\end{equation}
In Eq. \eqref{eq:model_vectorised}, the number of parameters of the $n$-dimensional TVP-VAR($1$) model  is $(T-1)n^2 + n(n+1)/2 = O(n^2)$, thus scales quadratically in $n$. In macroeconomic and financial applications, the number of variables of interest ranges from $n=3$ (small size model) to $n=20$ (large model) and even $n=100$ (huge model). This highlights the twofold need for shrinkage estimation methods and in particular, for the introduction of sparse estimation of the coefficient matrix. In fact, it is very hard both to provide a meaningful interpretation for a large VAR with full time-varying matrix $\mathbf{B}_t$ and to have an efficient and computationally feasible algorithm for an unrestricted estimation.% moreover, such an unrestricted estimation would be computationally daunting, if not impossible (in small $T$ samples).

Motivated by this fact, we provide a prior distribution, which allows for sparse estimation in a time-varying parameter setting. For each coefficient of the matrix of parameters $B_t$, we introduce a mixture prior with independent location and scale parameters:
\begin{equation*}
P(\boldsymbol{\beta}_t) = \prod_{j=1}^{n^2} P(\beta_{j,t}), \qquad t=2,\ldots,T.
%P(\boldsymbol{\beta}_t) = \prod_{j=1}^{n^2} \mathcal{D}E(\beta_{j,t} | \mu_{j,t}, \tau_{j,t}), \qquad t=2,\ldots,T.
\end{equation*}
where $P(\beta_{j,t})$ is the probability distribution of the vector matrix of coefficients (for example, we can choose it as a Double Exponential or Laplace distribution).
One of the most successful and widespread approach in the Bayesian literature consists in the use of (independent) spike-and-slab prior distributions (e.g., \cite{Mitchell88SpikeSlab_priors}, \cite{George93VariableSelection}, \cite{Smith96BNP_SpikeSlab_Dirac},  \cite{George97SpikeSlabPrior}) for each coefficient $\beta_{j,t}$.
Based on it, we specify a spike-and-slab prior distribution for each $\beta_{j,t}$, with $j=1,\ldots,n^2$ and $t=2,\ldots,T$, of the form
\begin{equation}
\beta_{j,t} \sim \pi_t R(\beta_{j,t}) + (1-\pi_t)Q(\beta_{j,t}),
\label{eq:spike_slab}
\end{equation}
where $R,Q$ correspond to the spike and slab distributions, respectively, and $\pi_t$ is the time-varying mixing probability (i.e., the prior probability of the spike component). In the literature, we have two commonly choices for $D$:  a Dirac mass at $0$ such that $R(\beta_{j,t}) = \delta_{\lbrace 0 \rbrace}(\beta_{j,t})$; and a centered (in zero) Normal distribution $R(\beta_{j,t}) = \mathcal{N}(\beta_{j,t}|0,\tau_0)$.
The Dirac spike is a degenerate distribution that allows for variable selection as a by-product of the estimation. Instead, the choice of a continuous, diffuse prior (like a Gaussian) allows for shrinkage of the coefficients and is computationally faster, but requires the post-processing specification of threshold for the sake of variable selection.

The standard choice for the slab component $Q$ is a heavy-tailed distribution belonging to the family of Generalised Hyperbolic distribution (e.g., Double Exponential, Cauchy, $t$-Student), since the aim of this component is to capture potentially large non-zero coefficients. In the case of Dirac spike, the prior for each coefficient, for $j=1,\ldots,n^2$ and $t=2,\ldots,T$, is given by
\begin{align}
\beta_{j,t}|\mu_{j},\lambda_{j},\pi_t & \sim \pi_t \delta_{(0)}(\beta_{j,t}) + (1-\pi_t)\mathcal{N}(\beta_{j,t}|\mu_{j},\lambda_{j}),
\label{eq:prior_Dirac}
\end{align}
while for a diffuse (Gaussian) spike we have
\begin{align}
\label{eq:prior_diffuse_1}
\beta_{j,t}|\mu_{j},\lambda_{j},\pi_t,\tau_0 & \sim \pi_t \mathcal{N}(\beta_{j,t}|0,\tau_0) + (1-\pi_t)\mathcal{N}(\beta_{j,t}|\mu_{j},\lambda_{j}).
%, \\
%\label{eq:prior_diffuse_2}
%\tau_0 & \sim \mathcal{IG}(\tau_0|a_0,b_0).
\end{align}

\begin{example}
In \autoref{fig:prior_spike_slab} we report an example of spike-and-slab prior, with centred Gaussian spike distribution (in blue) and centred double exponential slab distribution (in red). From the left panel, which shows the two distributions, we can see that the Gaussian accounts for most of the prior mass on $0$ while the double exponential governs the tails. This is reflected in the plot on the right, which shows that the mixture distribution (with equal weights) has fatter tails than the Gaussian and more mass in a neighborhood of $0$ than the double exponential.

\begin{figure}[t!h]
\centering
\setlength{\abovecaptionskip}{1pt}
\begin{tabular}{cc}
\includegraphics[trim=5mm 0mm 10mm 0mm,clip, height= 4.3cm, width= 6.0cm]{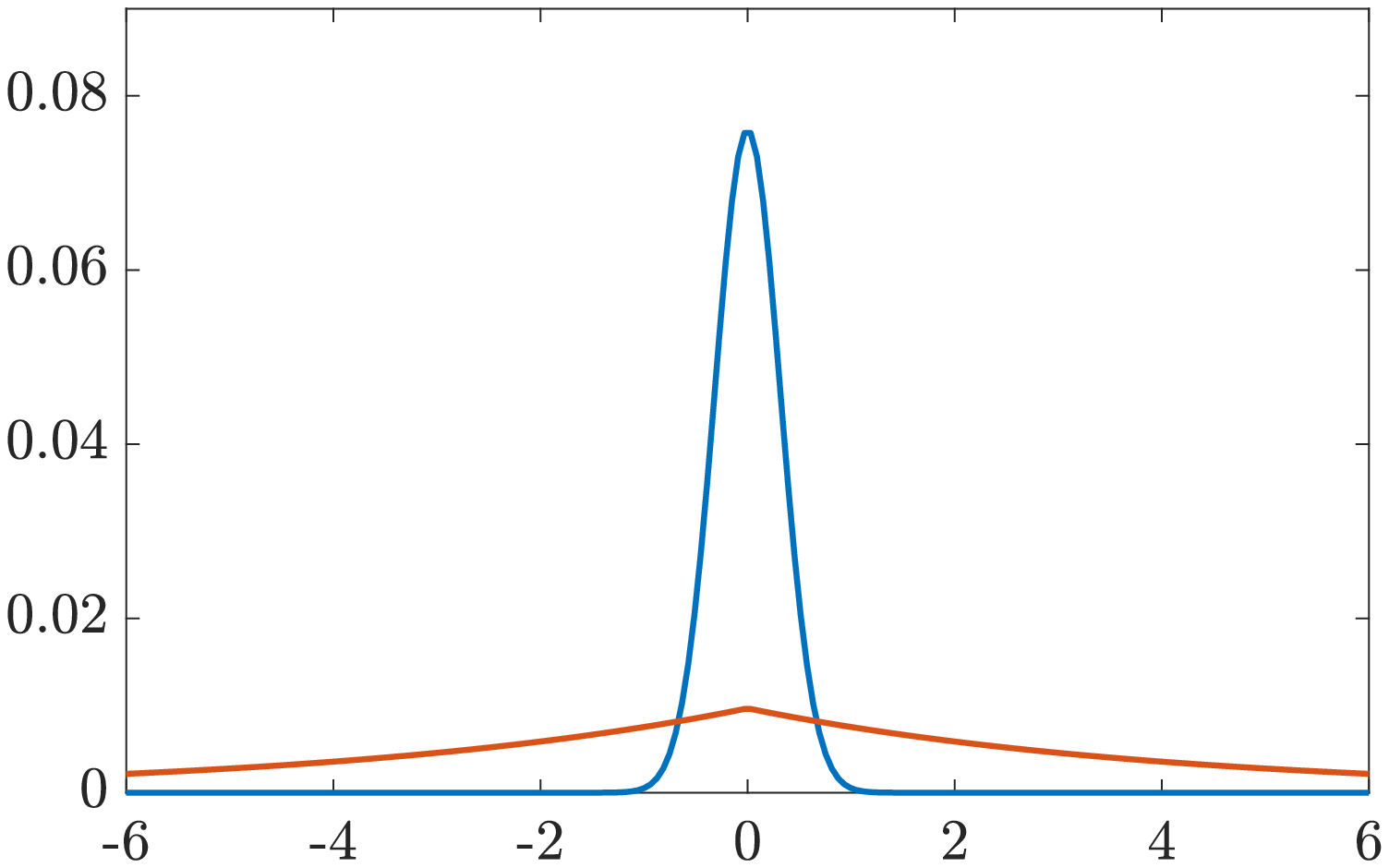} & 
\includegraphics[trim=5mm 0mm 10mm 0mm,clip, height= 4.3cm, width= 6.0cm]{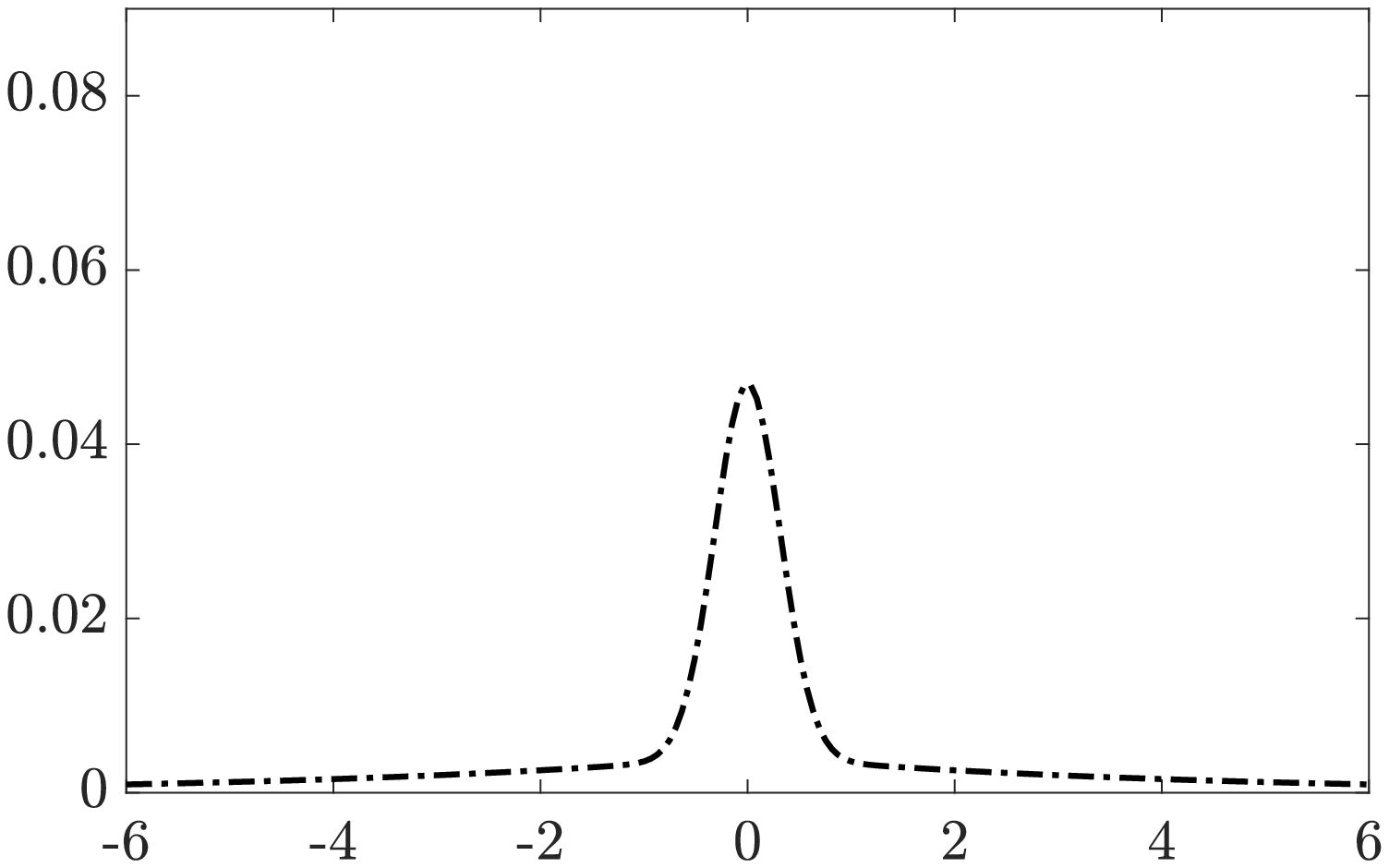}
\end{tabular}
\caption{Example of spike-and-slab distribution as in eq. \eqref{eq:prior_diffuse_1}. \textit{Left:} $\mathcal{N}(0,\sqrt{0.1})$ spike (\textit{blue}) and $\mathcal{D}E(0,4)$ slab (\textit{red}) components. \textit{Right:} mixture distribution, with mixing probability $\pi_t = 0.5$.}
\label{fig:prior_spike_slab}
\end{figure}
\end{example}

As previously described, we study the performance of the two different specification of the spike component by comparing the performances of the two constructions in extracting time-varying Granger causality networks from time series data.

The literature on time-varying parameter (TVP) models is vast. Some common parametric specifications include the threshold AR (TAR, e.g., \cite{Tong80ThresholdAR}), smooth transition AR (STAR, e.g.,\cite{Terasvirta94SmoothTransitionAR}), along with their multivariate generalisations, Markov switching process (e.g., \cite{Hamilton89MS}, \cite{Krolzig97MS}), change point process (e.g. \cite{Pesaran06ChangePoint_VAR}) and random walk process (e.g. \cite{DelNegroPrimiceri15Bayes_TVP_VAR}, \cite{Primiceri05Bayes_TVP_VAR}).
The choice of the particular specifications has been motivated by the intent to capture a particular feature of the dynamic evolution of the coefficients, such as changing regimes, structural breaks or smooth variations.

Differently from the existing literature on TVP-VAR models, we model the temporal dependence of the autoregressive parameters $\beta_{j,t}$ via assuming that the underlying prior (random) distributions evolve according to a discrete-time Markov process. A standard approach in Bayesian nonparametrics involves the specification of a Dirichlet Process (a.k.a. DP, see \cite{Ferguson73BNP_DirichletProcess}) or a Dirichlet Process mixture (a.k.a. DPM, see e.g., \cite{Lo84BNP_DirichletProcessMixture}) prior for the distribution of the parameters of interest. The use of DP and related priors for a random probability measure $P$ allows for clustering of the variables $x_i \distas{iid} P$.
%Moreover, performing exact posterior inference with DP/DPM priors has become feasible and fast thanks to the slice sampler approach proposed by \cite{Walker07SliceSampler_DPMixture}. 

%By exploiting the stick-breaking construction of \cite{Sethuraman94StcikBreaking_DP}, 

We proceed by introducing the prior temporal dependence between the random measures $P_2,\ldots,P_T$ via the time series Dirichlet Process (tsDDP) of \cite{NietoBarajas12TimeSeries_DDP}. As stated in the paper, for time series models, it is convenient to use dependence on the weights and common location. In practice, we apply a common discretization over the sequence of random measures, while the assumption of common weights and dependent location will lead to a discretization over the probability scale.
In opposite to \cite{Taddy10AR_DynamicSpatialPoisson}, who was working with equally spaced time points, we accommodate for unequal time points. In our analysis a latent binomial process to induce the desired correlation has been used, differently from the stick-breaking random probability measures as in \cite{Taddy10AR_DynamicSpatialPoisson}, which use a beta autoregression on the fractions of the stick-breaking constructions by mean of two sets of latent variables.

By exploiting the stick-breaking construction of \cite{Sethuraman94StcikBreaking_DP}, the time series Dirichlet Process imposes a dependence for the random probability measures
\begin{equation}
P_t(\cdot) = \sum_{i=1}^\infty w_{i,t} \delta_{(\boldsymbol{\theta}_{i,t})}(\cdot),
\end{equation}
where the locations are fixed $\boldsymbol{\theta}_{i,t} = \boldsymbol{\theta}_{i}$ and the weights $w_{i,t}$ vary over time. The dependence is described by a Markov process for each un-normalised stick-breaking weight $v_{i,t}$, with $i=1,\ldots,\infty$, via auxiliary variables $z_{i,t}$ (in the spirit of \cite{Pitt02AR_auxiliary_variables}, \cite{Pitt05AR_auxiliary_variables_general}), as follows
\begin{align}
v_{i,1} 				& \sim \mathcal{B}e(1,\alpha) \notag \\
z_{i,t} | v_{i,t}   	& \sim \mathcal{B}in(m_{i,t},v_{i,t}) \label{joint_laten}\\
v_{i,t+1} | z_{i,t} 	& \sim \mathcal{B}e(1+z_{i,t},\alpha+m_{i,t}-z_{i,t}). \notag
\end{align}
The hyper-parameter $m_{i,t}$ tunes the strength of the dependence between $P_t$ and $P_{t+1}$, such that $m_{i,t} =0$ implies $P_t \perp P_{t+1}$, while $m_{i,t} \to \infty$ implies $P_t = P_{t+1}$ with probability 1 (see \cite{NietoBarajas12TimeSeries_DDP}). Note that this construction implies that at each time $t=2,\ldots,T$, the marginal distribution of each random measure is a Dirichlet Process, that is
\begin{equation*}
P_t \sim DP(\alpha,P_0).
\end{equation*}
with total mass parameter $\alpha$ and base measure $P_0$, such that the base measure defines the expectation and the mass parameter is interpreted as the precision parameter.

\begin{figure}[t!h]
\centering
\hspace*{-7ex}
\setlength{\abovecaptionskip}{1pt}
\begin{tabular}{ccccc}
$t=1$ & $t=2$ & $t=3$ \\
\includegraphics[trim=5mm 0mm 12mm 0mm,clip, height= 4.3cm, width= 6.0cm]{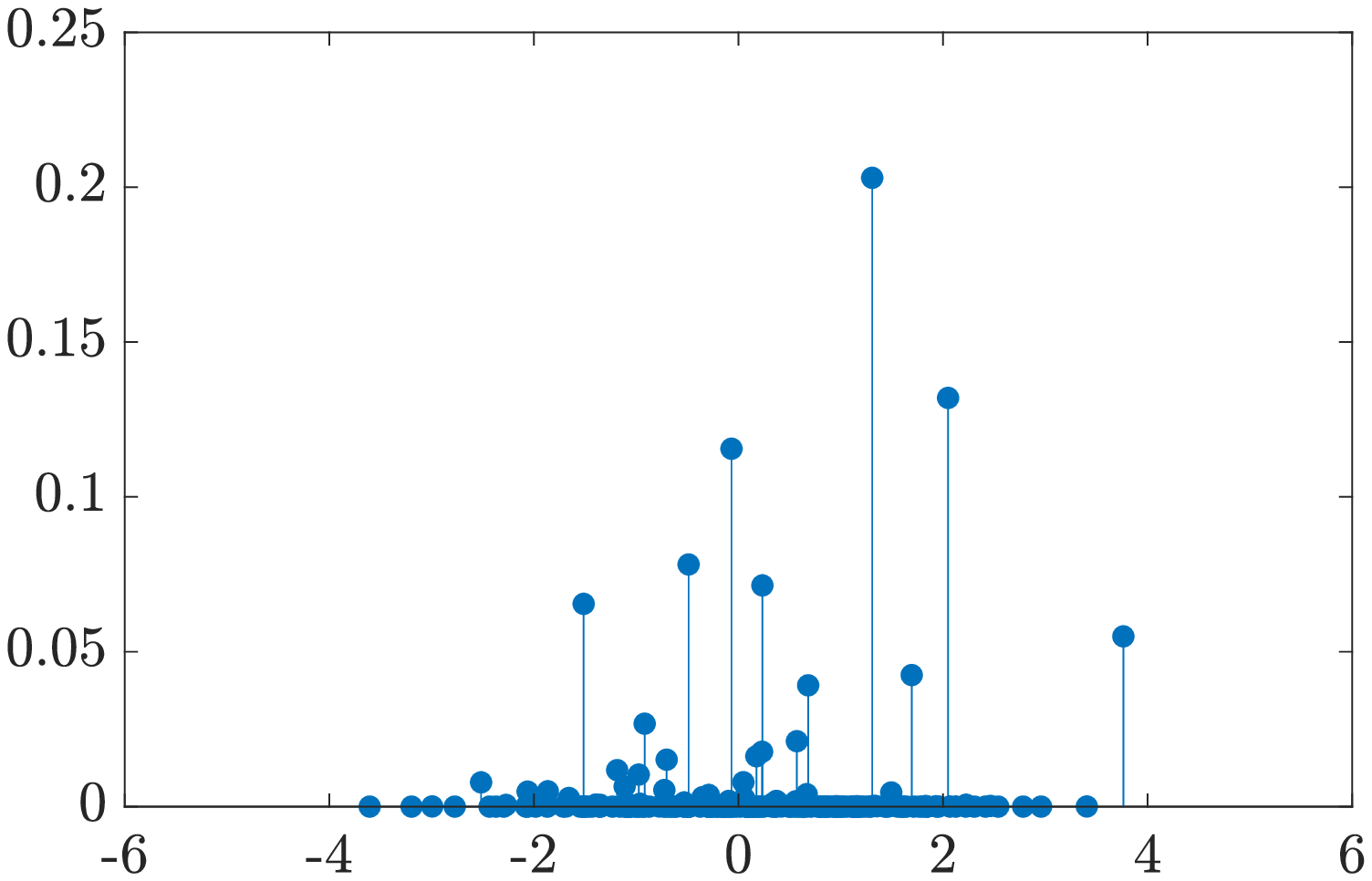} &
\includegraphics[trim=5mm 0mm 12mm 0mm,clip, height= 4.3cm, width= 6.0cm]{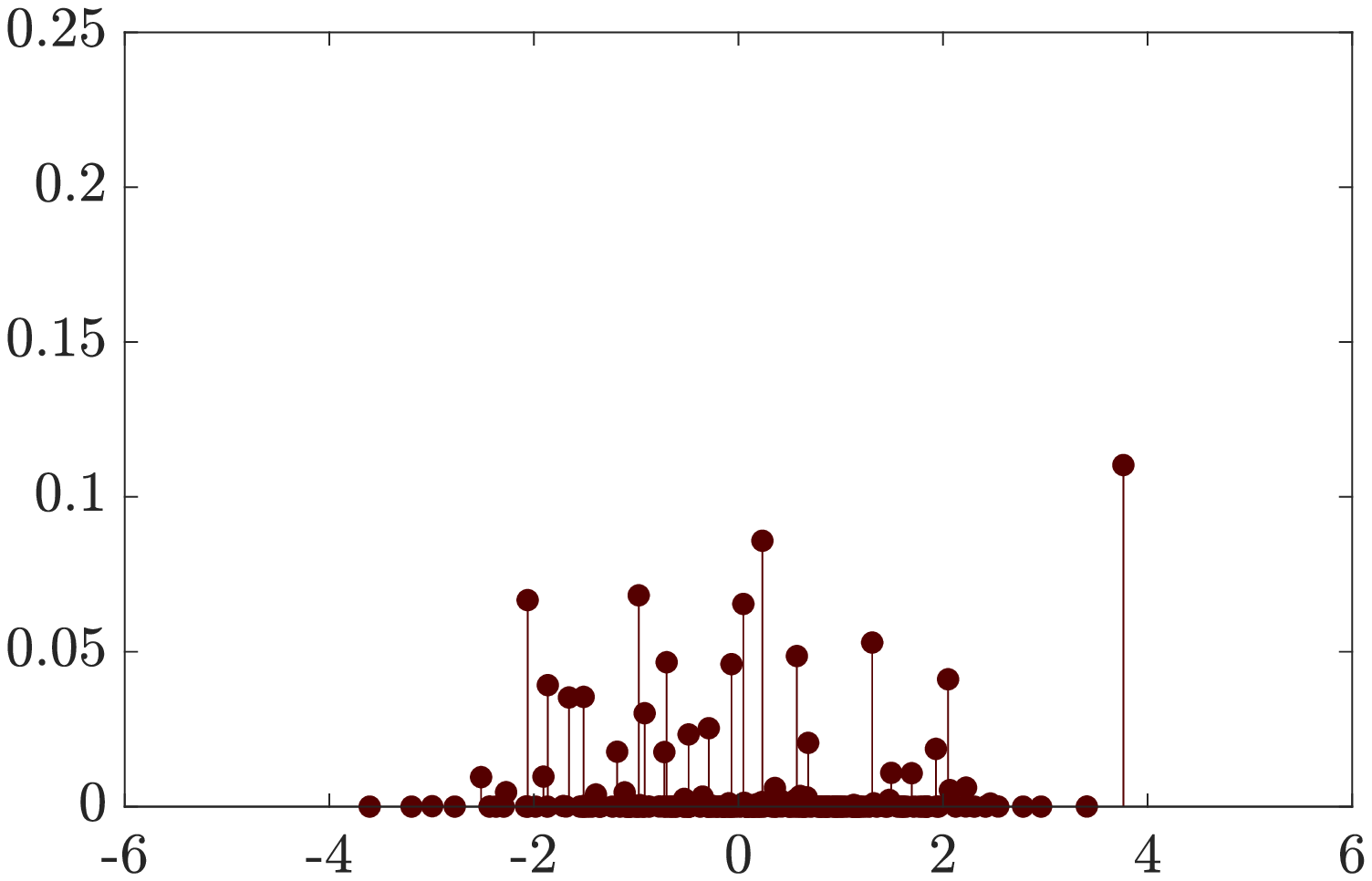} &
\includegraphics[trim=5mm 0mm 12mm 0mm,clip, height= 4.3cm, width= 6.0cm]{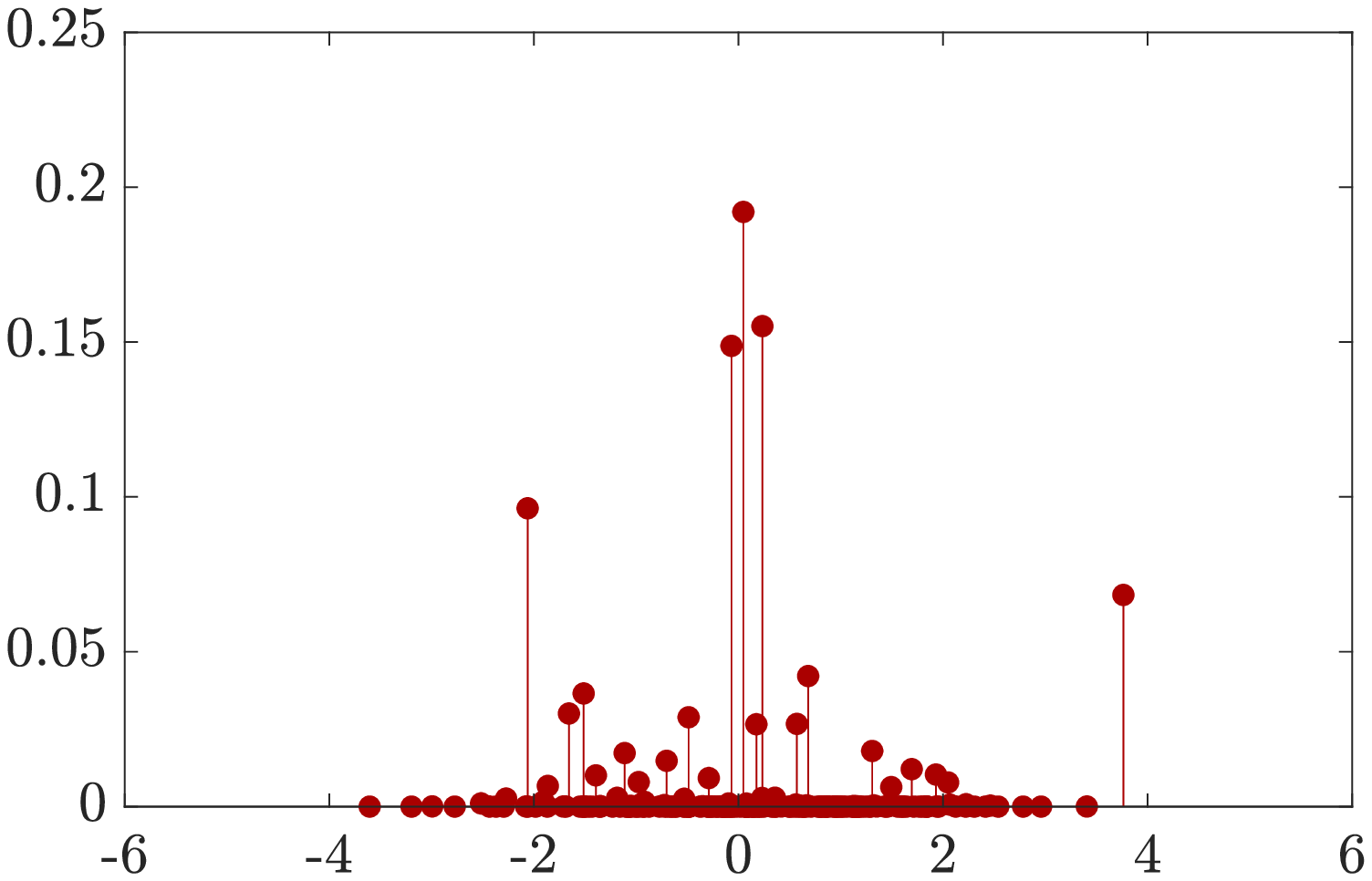} \\
 & & \\
 $t=4$ & $t=5$ & $t=6$\\
\includegraphics[trim=5mm 0mm 12mm 0mm,clip, height= 4.3cm, width= 6.0cm]{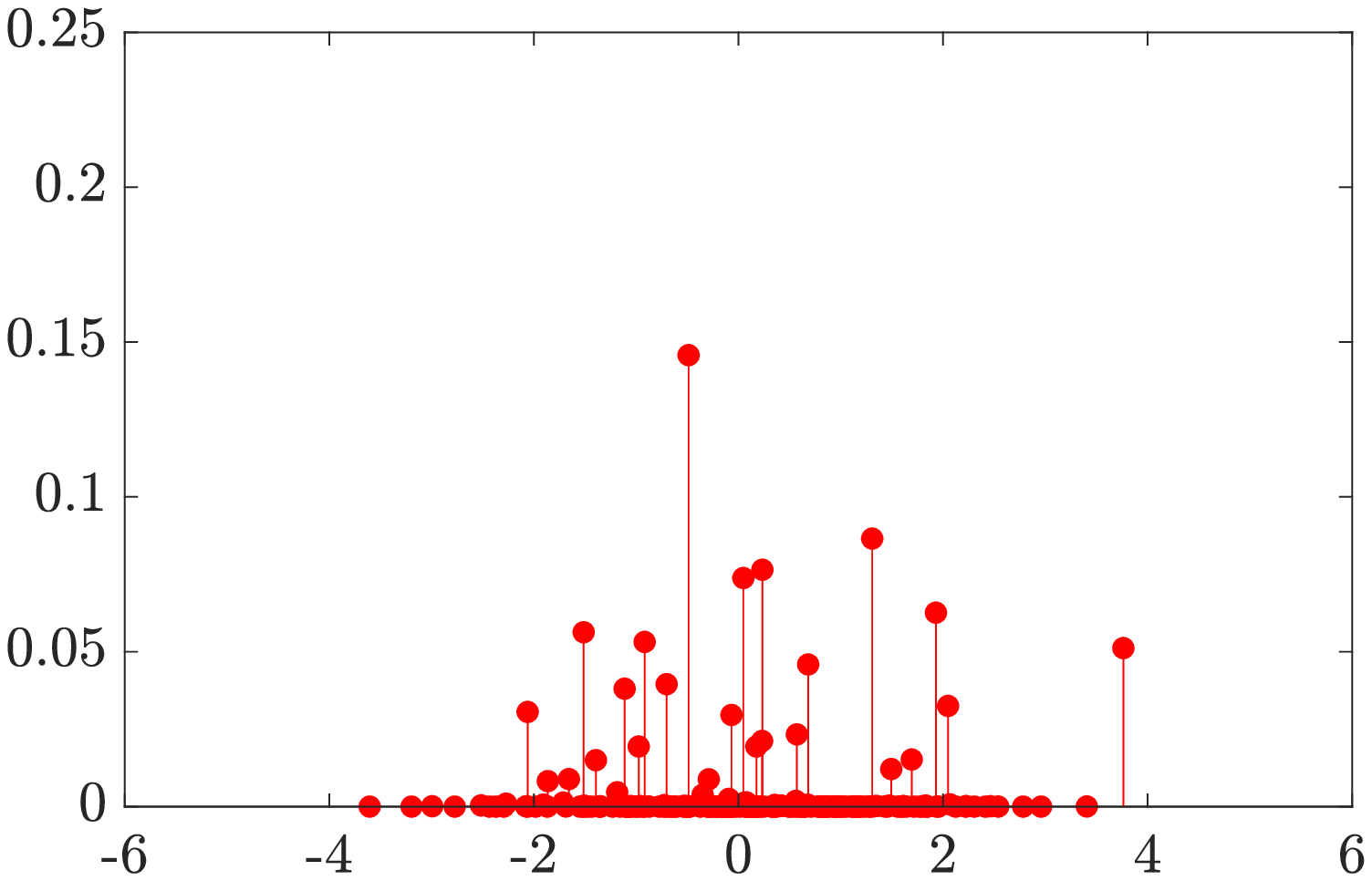} &
\includegraphics[trim=5mm 0mm 12mm 0mm,clip, height= 4.3cm, width= 6.0cm]{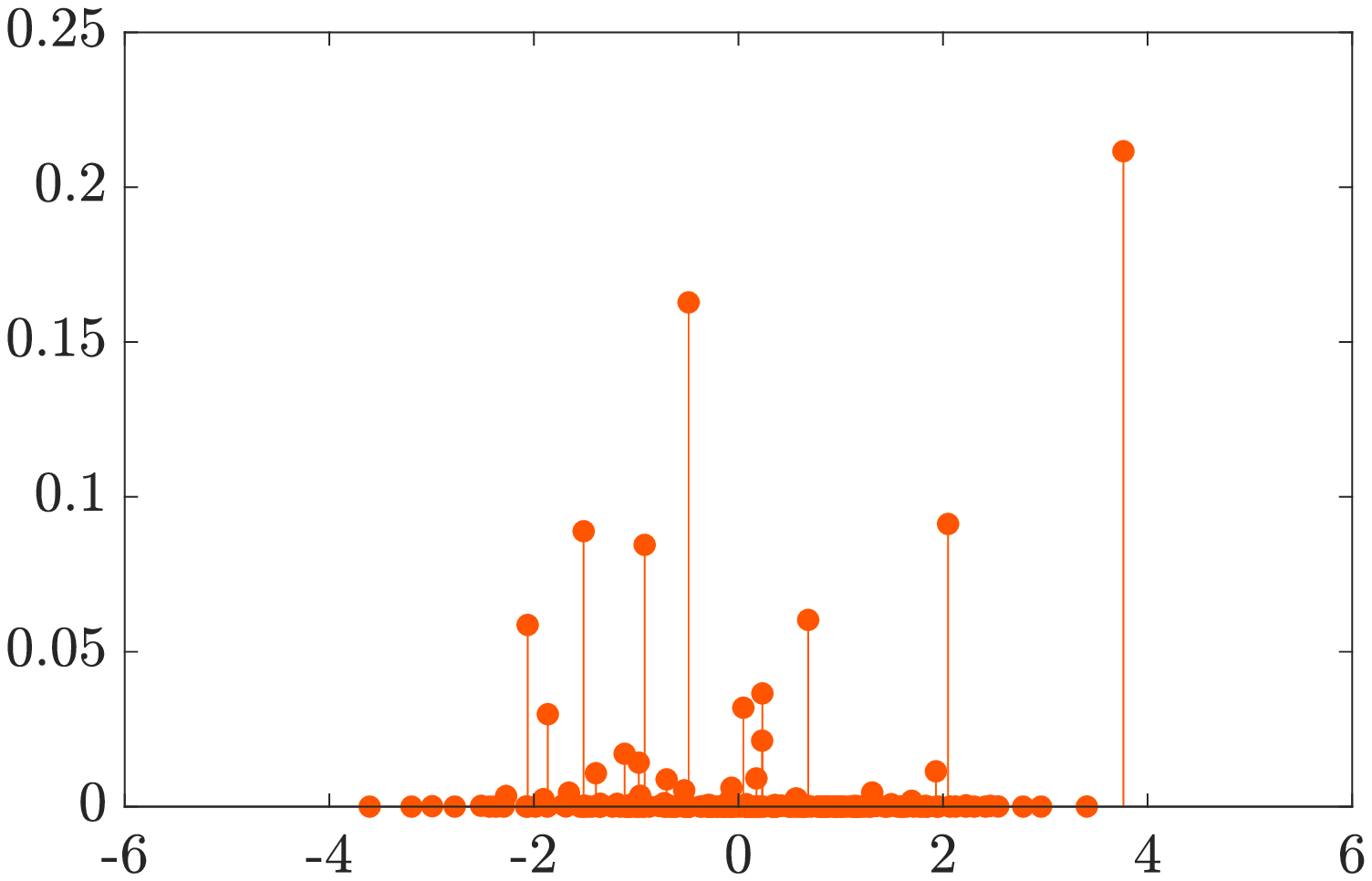} &
\includegraphics[trim=5mm 0mm 12mm 0mm,clip, height= 4.3cm, width= 6.0cm]{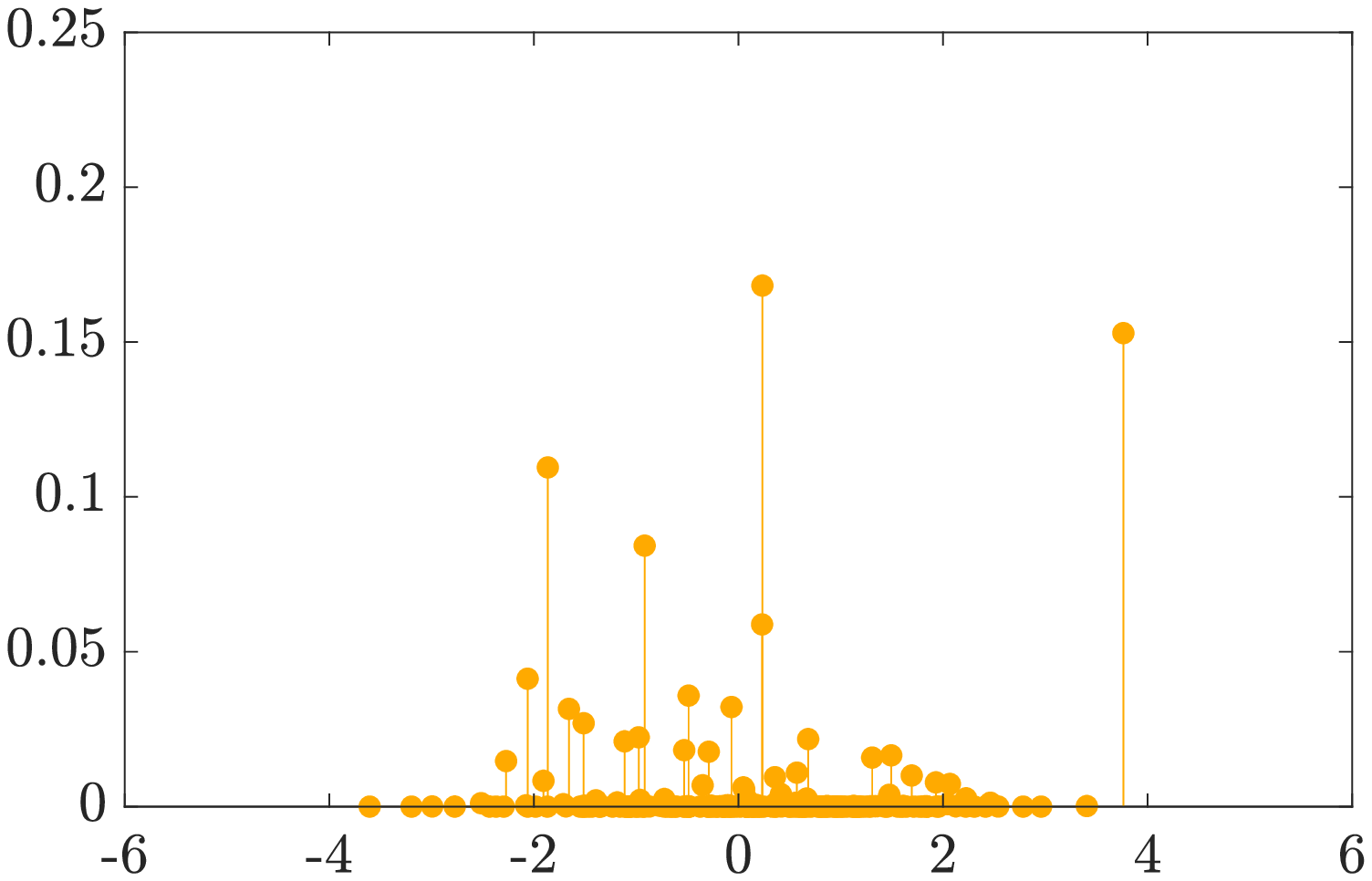} 
\end{tabular}
\caption{Sample from $tsDDP(\alpha,P_0,\mathbf{m})$, with base measure $P_0 \sim \mathcal{N}(0,\sqrt{2})$, concentration parameter $\alpha=10$ and strength of dependence $\mathbf{m}= m_{i,t}=4$. The size of the time series is assumed to be $T=6$.}
%\caption{Sample from $tsDDP(\alpha,P_0,\mathbf{m})$, with $P_0=\mathcal{N}(\mu_j|0,\sqrt{1.5})\mathcal{G}a(\tau_j|9,1/3)$, $\alpha=1$, $m_{i,t}=m=1$.}
\label{fig:prior_tsDDP}
\end{figure}

Eq. \eqref{joint_laten} explains the joint distribution of $z_{i,t}$ and $v_{i,t}$ and it allows us to define the joint model for $(P_1,\ldots,P_T)$ as a  $tsDDP(\alpha,P_0,\mathbf{m})$, where $\mathbf{m}$ is the sequence of the strength of dependence, $m_{i,t}$ for $i=1,2,\ldots$ and $t=1,\ldots,T$. In Figure \ref{fig:prior_tsDDP}, we show a single draw of $(P_1,\ldots,P_T) \sim tsDDP(\alpha,P_0,\mathbf{m})$, with total mass parameter $\alpha$ equal to $10$; base measure $P_0$ as a normal distribution with zero mean and variance $\sqrt{2}$, i.e. $P_0 \sim \mathcal{N}(0,\sqrt{2})$ and total timing $T=6$. 

In order to assess the dependence structure induced by the time series Dirichlet Process, we consider the correlation between two random probability measures $P_{t}$ and $P_{t+1}$. The following proposition is explaining this correlation:

\begin{proposition}[\cite{NietoBarajas12TimeSeries_DDP}]
Let $A \subset \mathbb{R}$ be measurable. For $(P_1,\ldots,P_T) \sim tsDDP(\alpha,P_0,\mathbf{m})$ and any $t=1,\ldots,T-1$ let $\rho_t(A) \coloneqq \textnormal{Corr}(P_t(A),P_{t+1}(A))$. Then
\begin{align*}
\rho_t(A) & = (1+\alpha) \sum_{h=1}^\infty a_{th} \prod_{i=1}^{h-1} b_{ti} + \frac{P_0(A)}{1-P_0(A)} \left[ \sum_{h=1}^\infty [2-(1+\alpha)a_{th}] \prod_{i=1}^{h-1} b_{ti} -(1+\alpha) \right],
\end{align*}
where
\begin{align*}
a_{ti} = \frac{2(1+m_{ti}) + \alpha}{(1+\alpha+m_{ti})(1+\alpha)(2+\alpha)}, \qquad b_{ti} = \frac{\alpha-1}{1+\alpha} + a_{ti}
\end{align*}
\end{proposition}

\begin{remark}
The correlation between $(P_t,P_{t+1})$ is larger in regions where the prior mean $P_0$ assigns more probability, meaning that the $tsDDP$ prior places strongest dependence in $P_0$-most probable regions. Note that strong dependence between $(P_t,P_{t+1})$ does not imply strong dependence between their outcomes $(\beta_{i,t},\beta_{j,t+1})$.
\end{remark}

We can summarize what we have described above in the following prior structure\footnote{We use the shape-scale parametrisation of the Gamma distribution (thus $\mathbb{E}[x] = ab$ and $\mathbb{V}[x] = ab^2$) and Inverse Gamma distribution, whose probability density functions are, respectively
\begin{align*}
x \sim \mathcal{G}a(x|a,b) & \iff p(x|a,b) = \frac{1}{b^a \Gamma(a)} x^{a-1} e^{-x/b}, \qquad x \in (0,\infty), \\
x \sim \mathcal{IG}(x|a,b) & \iff p(x|a,b) = \frac{b^a}{\Gamma(a)} x^{-a-1} e^{-b/x}, \qquad x \in (0,\infty).
\end{align*}
The exponential distribution is obtained as a particular case when $a=1$, that is $\mathcal{E}xp(x|b) = \mathcal{G}a(x|1,b)$.}, for $j=1,\ldots,n^2$ and $t=2,\ldots,T$ as follows
\begin{align}
\label{eq:prior_start}
\beta_{j,t}|\mu_j,\lambda_j,\pi_t & \sim \pi_t R(\beta_{j,t}) + (1-\pi_t)\mathcal{N}(\beta_{j,t}|\mu_j,\lambda_j) \\
\lambda_{j}|\tau_{j} & \sim \mathcal{E}xp(\lambda_{j}|2/\tau_{j}) \\
(\mu_{j},\tau_{j})|P_t & \sim P_t, \\
(P_2,\ldots,P_T) & \sim tsDDP(\alpha,P_0,\mathbf{m}) \\
P_0(\mu_{j},\tau_{j}) & \sim \mathcal{N}(\mu_{j} | c,d) \mathcal{G}a(\tau_{j} | a_1,b_1) \\
\pi_t & \sim \mathcal{B}e(1,\eta) 
\label{eq:prior_end}
\end{align}
where $R(\beta_{j,t})$ is either a Dirac mass at $0$ or a Normal distribution centered in zero and with variance $\tau_0$, which marginally has Inverse Gamma prior distribution $\tau_0 \sim \mathcal{IG}(\tau_0|a_0,b_0)$.
If we marginalize over $\lambda_{j}$, we have a Double Exponential slab distribution for each entry of the coefficient matrix. Following the notation of Eq. \eqref{eq:spike_slab} we have $Q(\beta_{j,t}) = \mathcal{D}E(\beta_{j,t}|\mu_{j},\tau_{j})$, resulting in
\begin{equation*}
\beta_{j,t}|\mu_{j},\tau_{j},\pi_t \sim \pi_t R(\beta_{j,t}) + (1-\pi_t)\mathcal{D}E(\beta_{j,t}|\mu_{j},\tau_{j}).
\end{equation*}
For the covariance matrix $\Sigma$, we assume an Inverse Wishart prior distribution that is:
\begin{equation}
\boldsymbol{\Sigma}  \sim \mathcal{IW}(\nu,\boldsymbol{\Psi})
\end{equation}
where $\nu$ and $\boldsymbol{\Psi}$ are the degrees of freedom and scale hyperparameters, respectively.

In summary, the observational model in Eq. \eqref{eq:model_general_vectorised} together with the prior structure in Eqs. \eqref{eq:prior_start} to \eqref{eq:prior_end} lead to the BNP-TVP-VAR(1) model. Eq. \eqref{eq:prior_start}--\eqref{eq:prior_end} represent our hierarchical prior and in Figure \ref{fig:flow_shrink_prior} we represent them through a Directed Acyclic Graph (DAG) for the Normal spike specification. The observable and non-observable random variables are indicated through shadow and empty circles, respectively.  On the left side we have the prior for $\Sigma$, while on the right side, we have the hierarchial prior for $\beta_t$, with a description of the first and second stage of the hierarchy by means of the shrinking parameters $\mu$, $\tau$ and $\lambda$.

\tikzstyle{hyper}= [circle, fill=white, inner sep=1pt, dashed, minimum size=20pt, font=\fontsize{10}{10}\selectfont, draw=black] %draw=black,
\tikzstyle{vertex} = [draw, black, circle, minimum size=25pt, inner sep=0pt]
\tikzstyle{plate} = [draw, rectangle, rounded corners, minimum width=34pt, minimum height=34pt, fit=#1]
\tikzstyle{dots} = [circle,text width=1.7em,minimum size=23pt,text centered, inner sep=0pt]
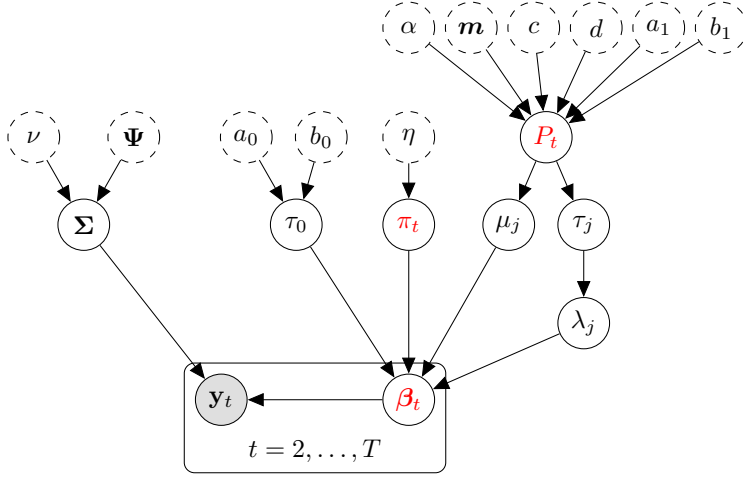
\begin{figure}
\setlength{\abovecaptionskip}{7pt}
\centering
\resizebox{10.0cm}{!}{
\begin{tikzpicture}[x=1.7cm,y=1.5cm] % node distance = 1.5cm, auto
    % Place nodes
	\node [hyper] (alpha) at (-6.0,5.2) {$\alpha$};
	\node [hyper] (m)  at (-5.5,5.2) {$\boldsymbol{m}$};
	\node [hyper] (c)  at (-5.0,5.2) {$c$};
	\node [hyper] (d)  at (-4.5,5.2) {$d$};
	\node [hyper] (a1) at (-4.0,5.2) {$a_1$};
	\node [hyper] (b1) at (-3.5,5.2) {$b_1$};
	\node [latent] (Pt) at (-4.9,4.2) {\textcolor{red}{$P_t$}};
    \node [hyper] (eta) at (-6.0,4.2) {$\eta$};
    	\node [hyper] (a0)  at (-7.3,4.2) {$a_0$};
    	\node [hyper] (b0)  at (-6.7,4.2) {$b_0$};
	\node [latent] (pi)   at (-6.0,3.4) {\textcolor{red}{$\pi_t$}};
    	\node [latent] (tau0) at (-6.9,3.4) {$\tau_0$};
    \node [latent] (muj)  at (-5.2,3.4) {$\mu_j$};
    \node [latent] (tauj) at (-4.6,3.4) {$\tau_j$};
    \node [latent] (lambdaj) at (-4.6,2.5) {$\lambda_j$};
    \node [latent] (beta) at (-6.0,1.8) {\textcolor{red}{$\boldsymbol{\beta}_t$}};
    	\node [hyper] (nu) at (-9.0,4.2) {$\nu$};
    \node [hyper] (Psi) at (-8.2,4.2) {$\boldsymbol{\Psi}$};
    	\node [latent] (Sigma) at (-8.6,3.4) {$\boldsymbol{\Sigma}$};
    	\node [obs] (Y) at (-7.5,1.8) {$\mathbf{y}_t$};
    % Draw edges
	\edge[] {alpha} {Pt};
	\edge[] {m} {Pt};
	\edge[] {c} {Pt};
	\edge[] {d} {Pt};
	\edge[] {a1} {Pt};
	\edge[] {b1} {Pt};
	\edge[] {Pt} {muj};
	\edge[] {Pt} {tauj};
    \edge[] {eta} {pi};
	\edge[] {a0} {tau0};
	\edge[] {b0} {tau0};
    \edge[] {muj} {beta};
    \edge[] {tauj} {lambdaj};
    \edge[] {lambdaj} {beta};
	\edge[] {tau0} {beta};
	\edge[] {pi} {beta};
    \edge[] {Psi} {Sigma};
    \edge[] {nu} {Sigma};
    \edge[] {Sigma} {Y};
    \edge[] {beta} {Y};
    % Plates
    \tikzset{plate caption/.append style={below =5pt and -10pt of #1.south}}
    \plate[] {plate l} {(Y) (beta)} {$t=2,\ldots,T$};
\end{tikzpicture}
}
\caption{\footnotesize DAG of the BNP-TVP-VAR model, with Normal spike. It exhibits the hierarchical structure of priors and related hyperparameters. The directed arrows show the causal dependence relations of the model.}
\label{fig:flow_shrink_prior}
\end{figure}

\vspace*{2ex}
Sufficient conditions for stationarity of TVP autoregressive models are given for the univariate case (despite the proof is valid also in the multivariate setting) in \cite{Brandt86Stationary_TVP_AR}, while \cite{Bourgerol92Stationary_TVP_AR_iid} provides conditions for multivariate regressions where the coefficients are independent and identically distributed. The sufficient conditions given by \cite{Brandt86Stationary_TVP_AR} is reported below.
\begin{theorem}[\cite{Brandt86Stationary_TVP_AR}]
Let $\lbrace (\mathbf{B}_t,\boldsymbol{\epsilon}_t), \; t \in \mathbb{Z} \rbrace$ be a strictly stationary ergodic process such that both $\mathbb{E}[\log^+(\norm{\mathbf{B}_0})]$ and $\mathbb{E}[\log^+(\norm{\boldsymbol{\epsilon}_0})]$ are finite. Suppose that the top Lyapunov exponent $\gamma$ defined by
\begin{equation*}
\gamma \coloneqq \inf_{t \in \mathbb{N}} \mathbb{E}\left[ \frac{1}{t+1} \log(\norm{\mathbf{B}_0 \mathbf{B}_{-1} \cdots \mathbf{B}_{-t}}) \right]
\end{equation*}
is strictly negative. Then, for all $t \in \mathbb{Z}$, the series
\begin{equation*}
\mathbf{y}_t = \sum_{i=0}^\infty \mathbf{B}_0 \mathbf{B}_{t-1} \cdots \mathbf{B}_{t-i+1} \boldsymbol{\epsilon}_{t-i}
\end{equation*}
converges a.s., and the process $\lbrace \mathbf{y}_t, \; t \in \mathbb{Z} \rbrace$ is the unique strictly stationary solution of
\begin{equation*}
\mathbf{y}_{t+1} = \mathbf{B}_{t+1} \mathbf{y}_t + \boldsymbol{\epsilon}_{t+1}, \qquad t \in \mathbb{Z}.
\end{equation*}
\end{theorem}

\subsection{Hyper-parameter elicitation}
Following \cite{NietoBarajas12TimeSeries_DDP}, we assume $m_{i,t} = m$, for each $i=1,2,\ldots$ and $t=1,\ldots,T$. Higher values of $m$ strengthen the dependence between the un-normalised weights $v_{i,t}$, however when big $m$ may induce the prior to overcome the likelihood, especially when the sample size is small. For this reason they specify a Poisson prior distribution for $m$, truncated on $\lbrace 1,\ldots,5 \rbrace$. Given the complexity of our prior specification, we prefer to fix the value of $m=5$, which is sufficiently small to avoid overweighting of the prior\footnote{In our empirical application, the sample size is $T=248$, while \cite{NietoBarajas12TimeSeries_DDP} have $T=8$.} and then check the robustness of the results to alternative values of $m$.
We choose the following values for the hyper-parameters:
\begin{equation*}
\begin{array}{cccccc}
c= 0 & d= 4.0 & a_1= 20.0 & b_1= 0.1 & a_0= 0.64 & b_0= 1.25, \\
\alpha= 1.0 & \eta= 1.0 & \nu = n+12 & \boldsymbol{\Psi} = \mathbf{I}_n/n.
\end{array}
\end{equation*}
This choice amounts to assuming a uniform prior on each $\pi_t$ and a rather uninformative prior on the covariance matrix, $\Sigma$. The value of the concentration parameter $\alpha$ is set according to standard practice in Dirichlet Process literature. The hyper-parameters $(c,d,a_1,b_1)$ imply that for each new component of the Dirichlet Process the prior distribution of $\mu_j$ is centered at zero mean with medium-high variance, whereas the prior for $\tau_j$ has mean $2$. Instead, the values of $(a_0,b_0)$ imply that the prior variance of the (diffuse) spike distribution is $0.8$, reflecting that this component should account for coefficients $\beta_{j,t}$ not significantly different from zero.

\section{Posterior computation}  \label{sec:posterior}
\subsection{Sampling method}

Since the joint posterior distribution is not tractable and it is complex to be sample from, Bayesian estimator cannot be obtained analytically. In this paper, we rely on simulation based inference methods, and develop a Gibbs sampler algorithm for approximating the posterior distribution.

%We designed and implemented a Gibbs sampler for posterior inference, since the joint posterior distribution is too complex to be sampled from, but all full conditional posterior distributions are available in closed form.
In order to deal with the finite mixture provided by the spike-and-slab prior and the infinite mixture given by the DPM, we exploited a data augmentation approach. For each $j=1,\ldots,n^2$ and $t=2,\ldots,T$, we introduce two sets of allocation variables $\gamma_{j,t},d_{j,t}$; a set of stick-breaking variables, $\mathbf{v}_{t} = \{v_{i,t}:i=1,2,\ldots\}$; a set of auxiliary variables $z_{i,t}$ (for $i=1,2,\ldots$) and a set of slice variables, $\mathbf{u}_{t} = \{u_{j,t}:j=1,\ldots,n^2\}$. The allocation variables, $\gamma_{j,t}$, selects the spike component $R(\cdot)$, when $\gamma_{j,t}$ is equal to zero and the slab component, when it is equal to one. The second allocation variable, $d_{j,t}$, selects the component of the Dirichlet mixture to which each single coefficient $\beta_{j,t}$ is allocated to. The sequence of stick-breaking variables defines the mixture weights, whereas the slice variable, $u_{j,t}$, allows us to deal with the infinite mixture components by identifying a finite number of stick-breaking variables to be sampled and an upper bound for the allocation variables $d_{j,t}$.

Finally, we obtain the following joint posterior distribution
\begin{equation}
\begin{split}
P(\mathbf{B}, \mathbf{U}, \mathbf{V}, \mathbf{Z}, & \mathbf{D}, \Gamma, \boldsymbol{\pi}, \boldsymbol{\mu}, \boldsymbol{\tau}, \boldsymbol{\lambda}, \Sigma | \mathbf{Y}) \propto L(\mathbf{Y} | \mathbf{B}, \Sigma) \cdot P(\Sigma) \cdot \prod_{k=1}^{k^*} P(\mu_k) P(\tau_k) \\
 & \cdot \prod_{t=2}^T P(\pi_t) \prod_{j=1}^{n^2} P(\beta_{j,t}|\pi_t,\lambda_j,\boldsymbol{\mu}) P(\lambda_j|\boldsymbol{\tau}) P(d_{j,t}|\mathbf{v}_{t},\mathbf{u}_{t}) P(\gamma_{j,t}|\pi_t) \\
 &  \cdot \prod_{i=1} P(u_{i,t}|v_{i,t}) P(v_{i,t}|z_{i,t-1}) P(z_{i,t-1}|v_{i,t-1}),
\end{split}
\label{eq:joint_posterior}
\end{equation}
where $\mathbf{U}=\{u_{j,t} : j= 1,\ldots, n^2; \mbox{ and } t = 2,\ldots,T\}$ and $\mathbf{V}=\{v_{i,t}: i =1,2,\ldots \mbox{ and } t=2,\ldots,T\}$ are the collections of slice variables and stick-breaking components, respectively; $\mathbf{Z} = \{z_{i,t}: i =1,2,\ldots \mbox{ and } t=2,\ldots,T\}$ and $\boldsymbol{\lambda} = \{\lambda_j: j = 1,\ldots,n^2\}$ are the auxiliary and latent variables, respectively; $\mathbf{D} = \{d_{j,t}: j= 1,\ldots, n^2; \mbox{ and } t = 2,\ldots,T\}$ and $\Gamma=\{\gamma_{j,t}: j= 1,\ldots, n^2; \mbox{ and } t = 2,\ldots,T \}$ are the allocation variables; $(\bm{\mu},\bm{\tau})= \{(\mu_{k},\tau_k): k =1,\ldots, k^{\ast}\}$ are the atoms, where $k$ ranges from $1$ to the number $k^*$ of allocated DP components; $\mathbf{B} = \{\bm{\beta}_t: t= 2,\ldots, T\}$ is the vector of VAR coefficients and $\bm{\pi} = \{\pi_t: t=2,\ldots,T \}$ are the specific probabilities of shrinking coefficients to zero.  

We obtain random samples from the posterior distributions by Gibbs sampling. The Gibbs sampler is based on the algorithm of  \cite{Hatjispyros11Dependent_DPM} and on the slice sampler approach of  \cite{Walker07SliceSampler_DPMixture} and \cite{Kalli11SliceSampler_DPM} for estimating the weights and locations of each random measure $P_t$. For improving the mixing of the MCMC, we introduced some Hamiltonian Monte Carlo (see \cite{Neal11HamiltonianMC}) steps in spite of drawing from the full conditional posterior distribution. Hereafter, we show the iterative steps by using the conditional independence between variables, for $k=1,\ldots,k^{\ast}$, $j= 1,\ldots, n^2$, $i=1,2,\ldots$ and $t=2,\ldots,T$:
\begin{enumerate}[label=(\arabic*)]
\item the slice and stick-breaking variables $u_{j,t}$ and $v_{i,t}$ are updated along with the auxiliary variable $z_{i,t}$ given $\left[ d_{j,t}, \gamma_{j,t} \right]$;
%$\left[\mu_{j}, \tau_j,\beta_{j,t},\Sigma, \lambda_{j}, d_{j,t}, \gamma_{j,t}, \pi_t, y_{t}\right]$;
\item the latent scale variables $\lambda_j$ are updated given $\left[ \bm{\mu}, \bm{\tau}, (\beta_{j,t}, d_{j,t}, \gamma_{j,t})_t \right]$;
%$\left[\mu_{j}, \tau_j,\beta_{j,t},\Sigma, u_{j,t}, v_{j,t}, z_{j,t}, d_{j,t}, \gamma_{j,t}, \pi_t, y_{t}\right]$;
\item the parameters of the stick-breaking locations $(\mu_k,\tau_k)$ are updated given $\left[ \bm{\lambda}, \mathbf{B}, \mathbf{D}, \Gamma \right]$;
%$\left[\lambda_{j},\beta_{j,t},\Sigma, u_{j,t}, v_{j,t}, z_{j,t}, d_{j,t}, \gamma_{j,t}, \pi_t, y_{t}\right]$;
\item the allocation variables $d_{j,t}, \gamma_{j,t}$ are jointly updated given $\left[ \mu_k, \tau_k, \beta_{j,t}, u_{j,t}, v_{i,t}, \pi_t \right]$;
%$\left[ \mu_j, \tau_j,\beta_{j,t},\Sigma, u_{j,t}, v_{j,t}, z_{j,t}, \lambda_{j}, \pi_t, y_{t} \right]$;
\item the VAR coefficients $\boldsymbol{\beta}_{t}$ are jointly updated given $\left[ \bm{\mu}, \bm{\tau}, \bm{\lambda}, \Sigma, (d_{j,t}, \gamma_{j,t})_j, \mathbf{y}_{t} \right]$;
%$\left[ \mu_k, \tau_k, \lambda_{j},\Sigma, u_{j,t}, v_{j,t}, z_{j,t}, d_{j,t}, \gamma_{j,t}, \pi_t, y_{t} \right]$;
\item The covariance matrix $\Sigma$ is updated given $\left[ \mathbf{B}, \mathbf{Y} \right]$;
%$\left[ \mu_k, \tau_k, \beta_{j,t},\lambda_j, u_{j,t}, v_{i,t}, z_{i,t}, d_{j,t}, \gamma_{j,t}, \pi_t, \mathbf{y}_{t} \right]$;
\item the mixing probability $\pi_t$ of having sparse coefficients is updated given $\left[(\gamma_{j,t})_j \right]$.
%$\left[ \mu_k, \tau_k, \beta_{j,t}, \Sigma, u_{j,t}, v_{i,t}, z_{i,t}, d_{j,t}, \gamma_{j,t}, \lambda_j, \mathbf{y}_{t} \right]$.
\end{enumerate}
The detailed Gibbs sampler is described in   \autoref{sec:apdx_posterior_diffuse} and \autoref{sec:apdx_posterior_Dirac}.

\subsection{Graph extraction} \label{sec:graph_extraction}

Based on the Gibbs sampler previously described, we are able to extract time-varying Granger-causal graphs. In the literature, linkages and networks describing the relationships between variables of interest, such as macroeconomics and financial linkages (e.g. \cite{Billioetal12GrangerNet} and \cite{Barigozzi19NETS_network_estimation}) can be used to extract pairwise Granger causality. This approach is generating spurious causality effects and does not consider conditioning on variables of interest. The main problem relies on the high number of variables available relative to the number of data, thus it could lead to overparametrization and inefficiency in gauging the causal relationships. Our proposed prior can be used to extract the networks and pairwise Granger causality while reducing the overfitting and curse of dimensionality problems. Moreover, the introduction of our prior could lead to the extraction of edge-colored graphs, that allows us to identify stylized facts in financial or macroeconomics networks and to show the presence of communities, hubs and linkage heterogeneity. 

From the MCMC output of the time-varying coefficient matrix $\mathbf{B}_t$, we are able to extract time-varying Granger-causal graphs.
At each time $t=2,\ldots,T$, we use the posterior random partition induced by the nonparametric (slab) distribution to cluster the edges of the graph (i.e., the entries $\beta_{j,t}$, $j=1,\ldots,n^2$ of the vectorised coefficient matrix $\boldsymbol{\beta}_t$) into groups. 

Formally, a graph $G$ is a pair $(V,E)$, where $V$ is a set of nodes and $E$ is a set of nodes pairs, named links or edges. The nodes are labeled and a link/edge is identified by the pair of nodes it connects, $(i,j)$.
In particular, we have the existence of an edge if and only if the time-varying VAR coefficients of the variable $y_{i,t-1}$ in the equation of $y_{j,t}$ is not null. In our network analysis, we focus on the adjacency matrix constructed a posterior from the allocation variables and it allows to take both values between $0$ and $1$ if we apply a threshold, while if the values are allowed to vary between $0$ and $1$, we have a weighted graph.
The purpose is to estimate the most significant time-varying dependence interrelationships (in terms of Granger-causality) between the $n$ variables of interest.

\begin{example}
Consider the TVP-VAR(1) model in Eq. \eqref{eq:model_general_vectorised} and let $n=4$. Without loss of generality, focus on the coefficient matrices at three consecutive times $t-1,t$ and $t+1$. Suppose the posterior estimates of the coefficient matrices and allocation variables $d_{j,t}$, respectively,  are as follows
\begin{equation}
\begin{array}{cccc}
\mathbf{B}_{t-1} & = \begin{bmatrix}
0 & 0 & 0 & 0.8 \\  0 & 0 & 0.8 & 0.2 \\  0.8 & 0.2 & 0 & 0 \\  0 & -0.4 & 0 & 0
\end{bmatrix} & \quad
\mathbf{D}_{t-1} & = \begin{bmatrix}
0 & 0 & 0 & 2 \\  0 & 0 & 2 & 1 \\  2 & 1 & 0 & 0 \\  0 & 3 & 0 & 0
\end{bmatrix} \\ \\
\mathbf{B}_{t} & = \begin{bmatrix}
0 & 0 & 0 & 0.8 \\  0.2 & 0 & 0.8 & -0.4 \\  0.2 & 0 & 0 & 0.8 \\  0 & -0.4 & 0 & 0
\end{bmatrix} & \quad
\mathbf{D}_{t} & = \begin{bmatrix}
0 & 0 & 0 & 2 \\  1 & 0 & 2 & 3 \\  1 & 0 & 0 & 2 \\  0 & 3 & 0 & 0
\end{bmatrix} \\ \\
\mathbf{B}_{t+1} & = \begin{bmatrix}
0 & 0 & 0 & 0 \\  0.2 & 0 & 0.8 & -0.4 \\  0.8 & 0 & 0 & 0.8 \\  0.2 & 0.2 & 0 & 0
\end{bmatrix} & \quad
\mathbf{D}_{t+1} & = \begin{bmatrix}
0 & 0 & 0 & 0 \\  1 & 0 & 2 & 3 \\  2 & 0 & 0 & 2 \\  1 & 1 & 0 & 0
\end{bmatrix}.
\end{array}
\label{eq:example_graph}
\end{equation}
The corresponding Granger-causal graphs are given in \autoref{fig:example_graph}, where colours have been used to denote the cluster assignment encoded in the matrices $\mathbf{D}_{t-1},\mathbf{D}_t$ and $\mathbf{D}_{t+1}$.

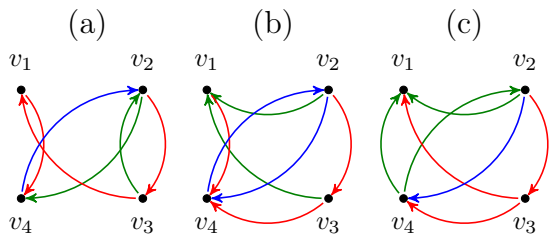
\begin{figure}[t!h]
\centering
\tikzstyle{empty}= [circle, fill=white, draw=white, inner sep=1pt, solid, minimum size=2pt, font=\fontsize{10}{10}\selectfont]
\tikzstyle{latent}= [circle, fill=black, draw=black, inner sep=1pt, solid, minimum size=2pt, font=\fontsize{10}{10}\selectfont]
\tikzstyle{Edge} = [very thick]
\setlength{\tabcolsep}{3pt}
\begin{tabular}{ccc}
(a) & (b) & (c)  \\
	%%%%%%%%% graph
	\begin{tikzpicture}[x=1.6cm,y=1.8cm]
	\node [latent] (v1) at (-6.5,3.4) {};
	\node [empty] (v1lab) at (-6.5,3.6) {$v_{1}$};
	\node [latent] (v2) at (-5.5,3.4) {};
	\node [empty] (v2lab) at (-5.5,3.6) {$v_{2}$};
	\node [latent] (v4) at (-6.5,2.6) {};
	\node [empty] (v4lab) at (-6.5,2.4) {$v_{4}$};
	\node [latent] (v3) at (-5.5,2.6) {};
	\node [empty] (v3lab) at (-5.5,2.4) {$v_{3}$};
	% Draw edges
	\tikzstyle{EdgeStyle}=[post]
	\tikzstyle{EdgeStyle}=[post,bend right=-40,darkgreen]
	\Edge[](v2)(v4)
	\Edge[](v3)(v2)
	\tikzstyle{EdgeStyle}=[post,bend right=-40,red]
	\Edge[](v3)(v1)
	\Edge[](v2)(v3)
	\Edge[](v1)(v4)
	\tikzstyle{EdgeStyle}=[post,bend right=-40,blue]
	\Edge[](v4)(v2)
	\end{tikzpicture}
	& %%%%%%%%% graph t
	\begin{tikzpicture}[x=1.6cm,y=1.8cm]
	\node [latent] (v1) at (-6.5,3.4) {};
	\node [empty] (v1lab) at (-6.5,3.6) {$v_{1}$};
	\node [latent] (v2) at (-5.5,3.4) {};
	\node [empty] (v2lab) at (-5.5,3.6) {$v_{2}$};
	\node [latent] (v4) at (-6.5,2.6) {};
	\node [empty] (v4lab) at (-6.5,2.4) {$v_{4}$};
	\node [latent] (v3) at (-5.5,2.6) {};
	\node [empty] (v3lab) at (-5.5,2.4) {$v_{3}$};
	% Draw edges
	\tikzstyle{EdgeStyle}=[post]
	\tikzstyle{EdgeStyle}=[post,bend right=-40,darkgreen]
	\Edge[](v2)(v1)
	\Edge[](v3)(v1)
	\tikzstyle{EdgeStyle}=[post,bend right=-40,red]
	\Edge[](v1)(v4)
	\Edge[](v2)(v3)
	\Edge[](v3)(v4)
	\tikzstyle{EdgeStyle}=[post,bend right=-40,blue]
	\Edge[](v2)(v4)
	\Edge[](v4)(v2)
	\end{tikzpicture}
	& %%%%%%%%% graph t+1
    \begin{tikzpicture}[x=1.6cm,y=1.8cm]
	\node [latent] (v1) at (-6.5,3.4) {};
	\node [empty] (v1lab) at (-6.5,3.6) {$v_{1}$};
	\node [latent] (v2) at (-5.5,3.4) {};
	\node [empty] (v2lab) at (-5.5,3.6) {$v_{2}$};
	\node [latent] (v4) at (-6.5,2.6) {};
	\node [empty] (v4lab) at (-6.5,2.4) {$v_{4}$};
	\node [latent] (v3) at (-5.5,2.6) {};
	\node [empty] (v3lab) at (-5.5,2.4) {$v_{3}$};
	% Draw edges
	\tikzstyle{EdgeStyle}=[post]
	\tikzstyle{EdgeStyle}=[post,bend right=-40,darkgreen]
	\Edge[](v2)(v1)
	\Edge[](v4)(v1)
	\Edge[](v4)(v2)
	\tikzstyle{EdgeStyle}=[post,bend right=-40,red]
	\Edge[](v2)(v3)
	\Edge[](v3)(v1)
	\Edge[](v3)(v4)
	\tikzstyle{EdgeStyle}=[post,bend right=-40,blue]
	\Edge[](v2)(v4)
	\end{tikzpicture}
\end{tabular}
\caption{Weighted graphs with $\hat{K}=3$ intensity levels: $\hat{\mu}_{1}^{\ast}= 0.2$ (green edges), $\hat{\mu}_{2}^{\ast}= 0.8$ (red edges) and $\hat{\mu}_{3}^{\ast}= -0.4$ (blue edges). In each graph the node $v_i$ represents the variable $i$ in the 4-dimensional VAR(1) in eq.~\eqref{eq:example_graph}, a clockwise-oriented edge from node $j$ to node $i$ represents a non-null coefficient for the variable $y_{j,t-1}$ in the $i$-th equation of the VAR. The vertex set is $V=\{v_1,v_2,v_3,v_4\}$ and the edges are $e_1=\{v_1,v_4\}$, $e_2=\{v_2,v_3\}$, $e_3=\{v_2,v_4\}$, $e_4=\{v_3,v_1\}$, $e_5=\{v_3,v_2\}$, $e_6=\{v_4,v_3\}$, $e_7=\{v_2,v_1\}$, $e_8=\{v_3,v_4\}$, $e_9=\{v_4,v_1\}$. Panel (a): weighted graph $G_{t-1}=(V,E_{t-1})$ with $E_{t-1}=\{e_1,e_2,e_3,e_4,e_5,e_6\}$ induced by edges of intensity level $\hat{\mu}_{1}^{\ast}= 0.2$.
Panel (b): the weighted graph $G_t=(V,E_t)$ with $E_t=\{e_1,e_2,e_3,e_4,e_6,e_7,e_8\}$ induced by edges of intensity level $\hat{\mu}_{2}^{\ast}= 0.8$.
Panel (c): the weighted graph $G_{t+1}=(V,E_{t+1})$ with $E_{t+1}=\{e_2,e_3,e_4,e_6,e_7,e_8,e_9\}$ induced by edges with intensity $\hat{\mu}_{3}^{\ast}= -0.4$.}
%\caption{Weighted graphs with $\hat{K}=2$ intensity levels: $\hat{\mu}_{1}^{\ast}=0.1$ (blue edges) and $\hat{\mu}_{2}^{\ast}=0.3$ (red edges). In each graph the node $v_i$ represents the variable $i$ in the 4-dimensional VAR(1), a clockwise-oriented edge from node $j$ to node $i$ represents a non-null coefficient for the variable $y_{j,t-1}$ in the $i$-th equation of the VAR. Panel (a): weighted graph $G=(V,E,C)$ (top) with vertex set $V=\{v_1,v_2,v_3,v_4\}$, edge set $E=\{e_1,e_2,e_3,e_4\}$, where $e_1=\{v_1,v_2\}$, $e_2=\{v_1,v_3\}$, $e_3=\{v_1,v_4\}$, $e_4=\{v_2,v_3\}$, $e_5=\{v_3,v_1\}$, $e_6=\{v_4,v_2\}$, $e_7=\{v_4,v_3\}$ and the adjacency matrix $A$ (middle) and the weights matrix $C$ (bottom).  Panel (b): the subgraph $G_1=(V,E_1)$ with $E_1=\{e_1,e_6,e_7\}$ induced by edges of intensity level $\hat{\mu}_{1}^{\ast}=0.3$. Panel (c): the subgraph $G_2=(V,E_2)$ with $E_2=\{e_2,e_3,e_4,e_5\}$ induced by edges with intensity $\hat{\mu}_{2}^{\ast}=0.3$. Panel (d): graph with two communities (red $V_1=\{v_1,v_2\}$ and blue $V_{2}=\{v_3,v_4\}$). Panel (e): graph with a hub (node $v_1$).}
\label{fig:example_graph}
\end{figure}

%\begin{figure}[t!h]
%\centering
%\setlength{\abovecaptionskip}{1pt}
%\begin{tabular}{ccc}
%$t-1$ & $t$ & $t+1$ \\
%\includegraphics[trim=10mm 0mm 10mm -15mm,clip, height= 4.4cm, width= 4.4cm]{ex_graph1.png} \quad & 
%\includegraphics[trim=10mm 0mm 10mm -15mm,clip, height= 4.4cm, width= 4.4cm]{ex_graph2.png} \quad &
%\includegraphics[trim=10mm 0mm 10mm 15mm,clip, height= 4.4cm, width= 4.4cm]{ex_graph3.png}
%\end{tabular}
%\caption{Extracted Granger causal graphs from the coefficient matrices in eq. \eqref{eq:example_graph}. Colours denote the cluster assignment encoded in $d_{j,t}$ (green is $1$, orange is $2$, cyan is $3$). No self-loops are displayed. Edges' clockwise orientation indicated the directionality of Granger causality.}
%\label{fig:example_graph}
%\end{figure}
\end{example}

\section{Conclusions} \label{sec:conclusions}
We proposed the BNP-TVP-VAR model for sparse, nonparametric inference in time-varying VAR models. The use of spike-and-slab priors with time-series dependent Dirichlet Process prior for the slab component allows to contemporaneously shrink the autoregressive coefficients and flexibly modelling time-varying non-zero entries.
We applied the proposed methodology using two alternative spike distributions: a Dirac and a Normal distribution. The performance of the resulting models has been compared in terms of: (i) sparse estimation and variable selection, and (ii) clustering structure.
Moreover, we showed how the BNP-TVP-VAR model can be used for extracting Granger-causal time-dependent graphs from multivariate time series.

%In the empirical application, we studied the macroeconomic dataset of \cite{McCracken16FRED-MD_dataset} with small, medium, and large size.

%%%%%%%%%%%%%%% BIBLIOGRAPHY %%%%%%%%%%%%%%%
\bibliographystyle{apalike}  % plain apalike siam chicago ecta
\bibliography{Biblio_BNP_TVP_VAR.bib}

\begin{thebibliography}{}

\bibitem[Barigozzi and Brownlees, 2019]{Barigozzi19NETS_network_estimation}
Barigozzi, M. and Brownlees, C. (2019).
\newblock Nets: Network estimation for time series.
\newblock {\em Journal of Applied Econometrics}, 34(3):347--364.

\bibitem[Bassetti et~al., 2014]{Bassetti14BetaPitmanYor}
Bassetti, F., Casarin, R., and Leisen, F. (2014).
\newblock Beta-product dependent {Pitman}--{Yor} processes for {Bayesian}
  inference.
\newblock {\em Journal of Econometrics}, 180(1):49--72.

\bibitem[Belmonte et~al., 2014]{Belmonte14shrink_TVP}
Belmonte, M. A.~G., Koop, G., and Korobilis, D. (2014).
\newblock Hierarchical shrinkage in time-varying parameter models.
\newblock {\em Journal of Forecasting}, 33(1):80--94.

\bibitem[Bianchi et~al., 2019]{Bianchi19GraphicalSUR}
Bianchi, D., Billio, M., Casarin, R., and Guidolin, M. (2019).
\newblock Modeling systemic risk with {Markov} switching graphical {SUR}
  models.
\newblock {\em Journal of Econometrics}, 2010(1):58--74.

\bibitem[Billio et~al., 2019]{Billio19BNP_sparseVAR}
Billio, M., Casarin, R., and Rossini, L. (2019).
\newblock Bayesian nonparametric sparse {VAR} models.
\newblock {\em Journal of Econometrics}, Forthcoming.

\bibitem[Billio et~al., 2012]{Billioetal12GrangerNet}
Billio, M., Getmansky, M., Lo, A.~W., and Pelizzon, L. (2012).
\newblock Econometric measures of connectedness and systemic risk in the
  finance and insurance sectors.
\newblock {\em Journal of Financial Economics}, 104(3):535--559.

\bibitem[Bitto and Fr{\"u}hwirth-Schnatter,
  2019]{Bitto19Shrinkage_Bayes_TVP_VAR}
Bitto, A. and Fr{\"u}hwirth-Schnatter, S. (2019).
\newblock Achieving shrinkage in a time-varying parameter model framework.
\newblock {\em Journal of Econometrics}, 210(1):75--97.

\bibitem[Bourgerol and Picard, 1992]{Bourgerol92Stationary_TVP_AR_iid}
Bourgerol, P. and Picard, N. (1992).
\newblock Strict stationarity of generalised autoregressive processes.
\newblock {\em Annals of Probability}, 20(1):1714--1730.

\bibitem[Brandt, 1986]{Brandt86Stationary_TVP_AR}
Brandt, A. (1986).
\newblock The stochastic equation {$Y_{n+1} = A_n Y_n+ B_n$} with stationary
  coefficients.
\newblock {\em Advances in Applied Probability}, 18(1):211--220.

\bibitem[Chan et~al., 2018]{Chan18CompLike_BayesVAR}
Chan, J., Eisenstat, E., Hou, C., and Koop, G. (2018).
\newblock Composite likelihood methods for large {Bayesian} {VAR}s with
  stochastic volatility.
\newblock {\em CAMA Working Paper}.

\bibitem[Chan et~al., 2016]{Chan16Large_BayesVARMA}
Chan, J., Eisenstat, E., and Koop, G. (2016).
\newblock Large {Bayesian} {VARMA}s.
\newblock {\em Journal of Econometrics}, 192(2):374--390.

\bibitem[Cogley and Sargent, 2005]{Cogley05Bayes_TVP_VAR_macro}
Cogley, T. and Sargent, T.~J. (2005).
\newblock Drifts and volatilities: monetary policies and outcomes in the post
  {WWII} {US}.
\newblock {\em Review of Economic Dynamics}, 8(2):262--302.

\bibitem[Creal et~al., 2013]{Creal13GAS}
Creal, D., Koopman, S.~J., and Lucas, A. (2013).
\newblock Generalized autoregressive score models with applications.
\newblock {\em Journal of Applied Econometrics}, 28(5):777--795.

\bibitem[Dangl and Halling, 2012]{Dangl12predictive_TVP}
Dangl, T. and Halling, M. (2012).
\newblock Predictive regressions with time-varying coefficients.
\newblock {\em Journal of Financial Economics}, 106(1):157--181.

\bibitem[Del~Negro and Primiceri, 2015]{DelNegroPrimiceri15Bayes_TVP_VAR}
Del~Negro, M. and Primiceri, G.~E. (2015).
\newblock Time varying structural vector autoregressions and monetary policy: A
  corrigendum.
\newblock {\em Review of Economic Studies}, 82(1342--1345).

\bibitem[Diebold and Yilmaz, 2009]{Diebold09VolatilitySpillover}
Diebold, F.~X. and Yilmaz, K. (2009).
\newblock Measuring financial asset return and volatility spillovers, with
  application to global equity markets.
\newblock {\em The Economic Journal}, 119(534):158--171.

\bibitem[Diebold and Yilmaz, 2012]{Diebold12GVAR_VolatilitySpillover}
Diebold, F.~X. and Yilmaz, K. (2012).
\newblock Better to give than to receive: Predictive directional measurement of
  volatility spillovers.
\newblock {\em International Journal of Forecasting}, 28(1):57--66.

\bibitem[Doan et~al., 1984]{DoanLitterman84BVAR_MinnesotaPrior_survey}
Doan, T., Litterman, R., and Sims, C. (1984).
\newblock Forecasting and conditional projection using realistic prior
  distributions.
\newblock {\em Econometric Reviews}, 3(1):1--100.

\bibitem[Ferguson, 1973]{Ferguson73BNP_DirichletProcess}
Ferguson, T.~S. (1973).
\newblock A {Bayesian} analysis of some nonparametric problems.
\newblock {\em The Annals of Statistics}, 1(2):209--230.

\bibitem[Gefang, 2014]{Gefang14Bayes_DoubleElasticNet}
Gefang, D. (2014).
\newblock Bayesian doubly adaptive elastic-net {Lasso} for var shrinkage.
\newblock {\em International Journal of Forecasting}, 30(1):1--11.

\bibitem[Gefang et~al., 2019]{Gefang19VariationalBayes_largeVAR}
Gefang, D., Koop, G., and Poon, A. (2019).
\newblock Variational bayesian inference in large vector autoregressions with
  hierarchical shrinkage.
\newblock {\em CAMA Working Paper}.

\bibitem[George and McCulloch, 1993]{George93VariableSelection}
George, E.~I. and McCulloch, R.~E. (1993).
\newblock Variable selection via gibbs sampling.
\newblock {\em Journal of the American Statistical Association},
  88(423):881--889.

\bibitem[George and McCulloch, 1997]{George97SpikeSlabPrior}
George, E.~I. and McCulloch, R.~E. (1997).
\newblock Approaches for {Bayesian} variable selection.
\newblock {\em Statistica Sinica}, 7:339--373.

\bibitem[Giannone et~al., 2014]{Giannone14priorVAR}
Giannone, D., Lenza, M., and Primiceri, G.~E. (2014).
\newblock Prior selection for vector autoregressions.
\newblock {\em The Review of Economics and Statistics}, 97(2):436--451.

\bibitem[Giannone et~al., 2018]{Giannone18IllusionSparsity}
Giannone, D., Lenza, M., and Primiceri, G.~E. (2018).
\newblock Economic predictions with big data: The illusion of sparsity.
\newblock {\em FRB of New York Staff Report No. 847. Available at SSRN:
  https://ssrn.com/abstract=3166281}.

\bibitem[Hamilton, 1989]{Hamilton89MS}
Hamilton, J.~D. (1989).
\newblock A new approach to the economic analysis of nonstationary time series
  and the business cycle.
\newblock {\em Econometrica}, 57(2):357--384.

\bibitem[Hatjispyros et~al., 2011]{Hatjispyros11Dependent_DPM}
Hatjispyros, S.~J., Nicoleris, T., and Walker, S.~G. (2011).
\newblock Dependent mixtures of dirichlet processes.
\newblock {\em Computational Statistics \& Data Analysis}, 55(6).

\bibitem[Huber and Feldkircher, 2019]{Huber19AdaptiveShrinkage_VAR}
Huber, F. and Feldkircher, M. (2019).
\newblock Adaptive shrinkage in {Bayesian} vector autoregressive models.
\newblock {\em Journal of Business \& Economic Statistics}, 37(1):27--39.

\bibitem[Kalli and Griffin, 2014]{Kalli14Bayes_sparse_TVP_VAR}
Kalli, M. and Griffin, J.~E. (2014).
\newblock Time-varying sparsity in dynamic regression models.
\newblock {\em Journal of Econometrics}, 178(2):779--793.

\bibitem[Kalli and Griffin, 2018]{Kalli18BNP_VAR}
Kalli, M. and Griffin, J.~E. (2018).
\newblock Bayesian nonparametric vector autoregressive models.
\newblock {\em Journal of Econometrics}, 203(2):267--282.

\bibitem[Kalli et~al., 2011]{Kalli11SliceSampler_DPM}
Kalli, M., Griffin, J.~E., and Walker, S.~G. (2011).
\newblock Slice sampling mixture models.
\newblock {\em Statistics and computing}, 21(1):93--105.

\bibitem[Karlsson, 2013]{Karlsson13Forecasting_BVAR_survey}
Karlsson, S. (2013).
\newblock Forecasting with bayesian vector autoregression.
\newblock In {\em Handbook of economic forecasting}, volume~2, pages 791--897.
  Elsevier.

\bibitem[Kastner and Huber, 2018]{Kastner18Sparse_largeVAR}
Kastner, G. and Huber, F. (2018).
\newblock Sparse {Bayesian} vector autoregressions in huge dimensions.
\newblock {\em arXiv preprint arXiv:1704.03239}.

\bibitem[Koop and Korobilis, 2010]{Koop10BVAR_survey}
Koop, G. and Korobilis, D. (2010).
\newblock Bayesian multivariate time series methods for empirical
  macroeconomics.
\newblock {\em Foundations and Trends in Econometrics}, 3(4):267--358.

\bibitem[Koop and Korobilis, 2013]{Koop13Large_TVP_VAR}
Koop, G. and Korobilis, D. (2013).
\newblock Large time-varying parameter vars.
\newblock {\em Journal of Econometrics}, 177(2):185--198.

\bibitem[Koop and Korobilis, 2018]{Koop18VariationalBayes_VAR}
Koop, G. and Korobilis, D. (2018).
\newblock Variational bayes inference in high-dimensional time-varying
  parameter models.
\newblock {\em arXiv preprint arXiv:1809.03031}.

\bibitem[Koop et~al., 2018]{KoopKorobilis18BayesianCompressedVAR}
Koop, G., Korobilis, D., and Pettenuzzo, D. (2018).
\newblock {Bayesian} compressed vector autoregressions.
\newblock {\em Journal of Econometrics}.

\bibitem[Korobilis, 2016]{Korobilis16PanelVAR_priorShrinkage}
Korobilis, D. (2016).
\newblock Prior selection for panel vector autoregressions.
\newblock {\em Computational Statistics \& Data Analysis}, 101:110--120.

\bibitem[Krolzig, 1997]{Krolzig97MS}
Krolzig, H.-M. (1997).
\newblock {\em {M}arkov-Switching Vector Autoregressions: Modelling,
  Statistical Inference, and Application to Business Cycle Analysis}.
\newblock Springer-Verlag Berlin Heidelberg.

\bibitem[Litterman, 1986]{Litterman86BVAR_MinnesotaPrior}
Litterman, R.~B. (1986).
\newblock Forecasting with bayesian vector autoregressions -- five years of
  experience.
\newblock {\em Journal of Business \& Economic Statistics}, 4(1):25--38.

\bibitem[Lo, 1984]{Lo84BNP_DirichletProcessMixture}
Lo, A.~Y. (1984).
\newblock On a class of {Bayesian} nonparametric estimates: I. density
  estimates.
\newblock {\em The Annals of Statistics}, 12(1):351--357.

\bibitem[McCracken and Ng, 2016]{McCracken16FRED-MD_dataset}
McCracken, M.~W. and Ng, S. (2016).
\newblock Fred-md: A monthly database for macroeconomic research.
\newblock {\em Journal of Business \& Economic Statistics}, 34(4):574--589.

\bibitem[Mitchell and Beauchamp, 1988]{Mitchell88SpikeSlab_priors}
Mitchell, T.~J. and Beauchamp, J.~J. (1988).
\newblock Bayesian variable selection in linear regression.
\newblock {\em Journal of the American Statistical Association},
  83(404):1023--1032.

\bibitem[Neal, 2011]{Neal11HamiltonianMC}
Neal, R.~M. (2011).
\newblock {MCMC} using {Hamiltonian} dynamics.
\newblock In Brooks, S., Gelman, A., Galin, J.~L., and Meng, X.-L., editors,
  {\em Handbook of {Markov Chain Monte Carlo}}, chapter~5. Chapman \& Hall
  /{CRC}.

\bibitem[Nieto-Barajas et~al., 2012]{NietoBarajas12TimeSeries_DDP}
Nieto-Barajas, L.~E., M{\:u}ller, P., Ji, Y., Lu, Y., and Mills, G.~B. (2012).
\newblock A {T}ime-{S}eries {DDP} for {F}unctional {P}roteomics {P}rofiles.
\newblock {\em Biometrics}, 68:859--868.

\bibitem[Pesaran et~al., 2006]{Pesaran06ChangePoint_VAR}
Pesaran, M.~H., Pettenuzzo, D., and Timmermann, A. (2006).
\newblock Forecasting time series subject to multiple structural breaks.
\newblock {\em The Review of Economic Studies}, 73(4):1057--1084.

\bibitem[Pitt et~al., 2002]{Pitt02AR_auxiliary_variables}
Pitt, M.~K., Chatfield, C., and Walker, S.~G. (2002).
\newblock Constructing first order stationary autoregressive models via latent
  processes.
\newblock {\em Scandinavian Journal of Statistics}, 29(4):657--663.

\bibitem[Pitt and Walker, 2005]{Pitt05AR_auxiliary_variables_general}
Pitt, M.~K. and Walker, S.~G. (2005).
\newblock Constructing stationary time series models using auxiliary variables
  with applications.
\newblock {\em Journal of the American Statistical Association},
  100(470):554--564.

\bibitem[Primiceri, 2005]{Primiceri05Bayes_TVP_VAR}
Primiceri, G.~E. (2005).
\newblock Time varying structural vector autoregressions and monetary policy.
\newblock {\em Review of Economic Studies}, 72(821--852).

\bibitem[Sethuraman, 1994]{Sethuraman94StcikBreaking_DP}
Sethuraman, J. (1994).
\newblock A constructive definition of dirichlet priors.
\newblock {\em Statistica sinica}, 4(2):639--650.

\bibitem[Sims, 1980]{Sims80VAR_macroeconomics}
Sims, C.~A. (1980).
\newblock Macroeconomics and reality.
\newblock {\em Econometrica}, 48(1):1--48.

\bibitem[Smith and Kohn, 1996]{Smith96BNP_SpikeSlab_Dirac}
Smith, M. and Kohn, R. (1996).
\newblock Nonparametric regression using {B}ayesian variable selection.
\newblock {\em Journal of Econometrics}, 75(2):317--343.

\bibitem[Stock and Watson, 2007]{Stock07US_inflation_forecast}
Stock, J.~H. and Watson, M.~W. (2007).
\newblock Why has {US} inflation become harder to forecast?
\newblock {\em Journal of Money, Credit and Banking}, 39(1):3--33.

\bibitem[Taddy, 2010]{Taddy10AR_DynamicSpatialPoisson}
Taddy, M.~A. (2010).
\newblock Autoregressive mixture models for dynamic spatial {Poisson}
  processes: Application to tracking intensity of violent crime.
\newblock {\em Journal of the American Statistical Association},
  105(492):1403--1417.

\bibitem[Ter{\"a}svirta, 1994]{Terasvirta94SmoothTransitionAR}
Ter{\"a}svirta, T. (1994).
\newblock Specification, estimation, and evaluation of smooth transition
  autoregressive models.
\newblock {\em Journal of the American Statistical Association},
  89(425):208--218.

\bibitem[Tong and Lim, 1980]{Tong80ThresholdAR}
Tong, H. and Lim, K. (1980).
\newblock Threshold autoregression, limit cycles and cyclical cata (with
  discussion of the paper).
\newblock {\em Journal of the Royal Statistical Society: Series {B}
  (Statistical Methodology)}, 42(3):245--292.

\bibitem[Walker, 2007]{Walker07SliceSampler_DPMixture}
Walker, S.~G. (2007).
\newblock Sampling the {D}irichlet mixture model with slices.
\newblock {\em Communications in Statistics---Simulation and Computation},
  36(1):45--54.

\end{thebibliography}

\clearpage

\appendix
\section{Posterior distributions: diffuse DE spike} \label{sec:apdx_posterior_diffuse}
\subsection{Posterior for stick breaking un-normalised weights $v_{i,t}$}
Posterior distribution for $v_{i,t}$, for all $t=1,\ldots,T$ and all $k=1,\ldots,k^*$, where $k^*$ is the number of ties. We use the convention $z_{i,0} = m_{i,0} = 0$, for all $i=1,2,\ldots$.
\begin{align*}
\notag
P(v_{k,t} | z_{k,t},z_{k,t-1}, \ldots) & \propto P(v_{k,t}|z_{k,t-1}) P(z_{k,t}|v_{k,t}) P(\ldots | v_{k,t}) \\ \notag
 & \propto v_{k,t}^{z_{k,t-1}} (1-v_{k,t})^{\alpha+m_{k,t-1}-z_{k,t-1}-1} \frac{m_{k,t}!}{z_{k,t}! (m_{k,t}-z_{k,t})!} v_{k,t}^{z_{k,t}} (1-v_{k,t})^{m_{k,t}-z_{k,t}} \\ \notag
 & \quad \cdot \left[ \prod_{j : d_{j,t} = k, \: \gamma_{j,t}=1} v_{j,t} \right] \left[ \prod_{j : d_{j,t} >k, \: \gamma_{j,t}=1} (1-v_{j,t}) \right] \\
 & \propto \mathcal{B}e(\overline{\xi}_t, \overline{\alpha}_t),
\end{align*}
where
\begin{align*}
\overline{\xi}_t & = 1+z_{k,t}+z_{k,t-1} + \sum_{j} \mathbb{I}(d_{j,t} = k, \gamma_{j,t} = 1), \\
\overline{\alpha}_t & = \alpha +m_{k,t}-z_{k,t} +m_{k,t-1}-z_{k,t-1} + \sum_{j} \mathbb{I}(d_{j,t} > k, \gamma_{j,t} = 1).
\end{align*}
The normalised weights are then computed by $w_{k,t} = v_{k,t} \prod_{j < k} (1-v_{j,t})$.
The posterior distribution of the latent auxiliary variables $z_{k,t}$, for all $t=1,\ldots,T-1$ and all $k=1,\ldots,k^*$ is given by
\begin{align*}
\notag
P(z_{k,t} | v_{k,t+1},v_{k,t}) & \propto P(z_{k,t} | v_{k,t}) P(v_{k,t+1} | z_{k,t}) \\ \notag
 & \propto \frac{m_{k,t}!}{z_{k,t}! (m_{k,t}-z_{k,t})!} v_{k,t}^{z_{k,t}} (1-v_{k,t})^{m_{k,t}-z_{k,t}} \\ \notag
 & \quad \cdot \frac{\Gamma(\alpha+m_{k,t}+1)}{\Gamma(z_{k,t}+1)\Gamma(\alpha+m_{k,t}-z_{k,t})} v_{k,t+1}^{z_{k,t}} (1-v_{k,t+1})^{\alpha+m_{k,t}-z_{k,t}-1} \\
 & \propto \frac{1}{(\Gamma(z_{k,t}+1))^2 \Gamma(\alpha+m_{k,t}-z_{k,t}) \Gamma(m_{k,t}-z_{k,t}+1)} \left(\frac{v_{k,t} v_{k,t+1}}{(1-v_{k,t})(1-v_{k,t+1})} \right)^{z_{k,t}}.
\end{align*}
Finally, the auxiliary variable for the slice sampler has posterior distribution given by
\begin{equation*}
P(u_{j,t}| \cdots) \propto \left\{
\begin{array}{ll}
\mathbb{I}(u_{j,t} < w_{d_{j,t}})^{\gamma_{j,t}}  	& \mathrm{if} \; \gamma_{j,t}=1 ,\\
\mathbb{I}(u_{j,t} < 1)^{1-\gamma_{j,t}} 			& \mathrm{if} \; \gamma_{j,t}=0.
\end{array} \right.
\end{equation*}

\subsection{Posterior for $\lambda_{j}$}
Posterior distribution for $\lambda_{j}$, for $j=1,\ldots,n^2$. Define $\mu_{k_j}^*,\tau_{k_j}^*$ are the location and scale of $\beta_{j,t}$, respectively, when sparse or non-sparse component is chosen.
\begin{align*}
\notag
P(\lambda_{j}&|\cdots)  \propto \prod_{t=2}^T \lambda_{j}^{-(1-\gamma_{j,t})/2} \exp\left\lbrace -(1-\gamma_{j,t}) \frac{1}{2\lambda_{j}} \beta_{j,t}^2 \right\rbrace \lambda_{j}^{-\gamma_{j,t}/2} \exp\left\lbrace -\gamma_{j,t} \frac{1}{2\lambda_{j}} (\beta_{j,t} -\mu_{d_{j,t}})^2 \right\rbrace \\ \notag
 & \quad \cdot \exp\left\lbrace (1-\gamma_{j,t}) \left( -\frac{\lambda_{j} \tau_0}{2} \right) \right\rbrace \exp\left\lbrace \gamma_{j,t} \left( -\frac{\lambda_{j} \tau_{d_{j,t}}}{2} \right) \right\rbrace \\ \notag
 & \propto \lambda_{j}^{-\frac{T-1}{2}} \exp\left\lbrace -\frac{1}{2}  \left[ \lambda_{j} \left( \sum_{t=2}^T (1-\gamma_{j,t}) \tau_0 + \gamma_{j,t} \tau_{d_{j,t}} \right) + \frac{1}{\lambda_j} \left( \sum_{t=2}^T (1-\gamma_{j,t})\beta_{j,t}^2 +\gamma_{j,t} (\beta_{j,t}-\mu_{d_{j,t}})^2 \right) \right] \right\rbrace \\
 & \propto GiG(p,a,b)
\end{align*}
where
\begin{equation*}
p=\frac{3-T}{2}, \qquad a= \sum_{t=2}^T (1-\gamma_{j,t}) \tau_0 + \gamma_{j,t} \tau_{d_{j,t}}, \qquad b= \sum_{t=2}^T (1-\gamma_{j,t})\beta_{j,t}^2 +\gamma_{j,t} (\beta_{j,t}-\mu_{d_{j,t}})^2.
\end{equation*}
The probability density function of a generalised inverse Gaussian, for $p\in \mathbb{R}, a >0, b >0$ and $K_p(\cdot)$ being a modified Bessel function of the second kind, is
\begin{equation*}
GiG(x|p,a,b) = \frac{(a/b)^{p/2}}{2 K_p(\sqrt{ab})} x^{p-1} e^{-(ax+b/x)/2}, \qquad x \in (0,\infty).
\end{equation*}
The vector $\boldsymbol{\lambda} = (\lambda_{1},\ldots,\lambda_{n^2})$ represents the diagonal of the diagonal matrix $\boldsymbol{\Lambda}$ of size $(n^2 \times n^2)$.

\subsection{Posterior of the stick breaking locations $(\mu_{k},\tau_{k})$}
The posterior distribution for the stick breaking locations in the sparse case are given by:
\begin{equation*}
\mu_{0} | \cdots \sim \delta_{(0)}(\mu_{0}).
\end{equation*}
and
\begin{align*}
\notag
P(\tau_{0} | \cdots) & \propto \mathcal{G}a(\tau_{0} | a_0,b_0) \cdot \prod_{(j,t) : \gamma_{j,t}=0} \frac{\tau_{0}}{2} \exp\left\lbrace -\frac{\tau_{0}}{2} \lambda_{j} \right\rbrace \\ \notag
 & \propto \tau_{0}^{a_0-1} \exp\left\lbrace -\frac{\tau_{0}}{b_0} \right\rbrace \cdot \prod_{(j,t) : \gamma_{j,t}=0} \frac{\tau_{0}}{2} \exp\left\lbrace -\frac{\tau_{0}}{2} \lambda_{j} \right\rbrace \\ \notag
 & \propto \tau_{0}^{a_0-1} \exp\left\lbrace -\frac{\tau_{0}}{b_0} \right\rbrace \tau_{0}^{m_0} \exp\left\lbrace -\frac{\tau_{0}}{2} \sum_{(j,t) : \gamma_{j,t}=0} \lambda_{j} \right\rbrace \\ \notag
 & \propto \tau_{0}^{a_0+m_0-1} \exp\left\lbrace -\tau_{0} \left( \frac{1}{b_0} + \frac{1}{2} \sum_{(j,t) : \gamma_{j,t}=0} \lambda_{j} \right) \right\rbrace \\
 & \propto \mathcal{G}a(\overline{a}_0, \overline{b}_0),
\end{align*}
with
\begin{equation*}
\overline{a}_0 = a_0 + m_0, \qquad \overline{b}_0 = \frac{2b_0}{2+b_0 \sum_{(j,t) : \gamma_{j,t}=0} \lambda_{j}},
\end{equation*}
where $m_0 = \sum_{(j,t)} \mathbb{I}(\gamma_{j,t}=0) = \sum_{j=1}^{n^2} \sum_{t=1}^T (1-\gamma_{j,t})$ and we used the parametrisation of the Gamma distribution with shape $a >0$ and scale $b >0$, that is
\begin{equation*}
\mathcal{G}a(x|a,b) = \frac{1}{b^a \Gamma(a)} x^{a-1} e^{-x/b}, \qquad x \in (0,\infty).
\end{equation*}

Regarding the sample in the non-sparse case, we generate $(\mu_k, \tau_k)$ for $k = 1,\ldots, N^{\ast}$ and we
have the following full conditional separately for $\mu_k$ and $\tau_k$.  The posterior distribution for the stick breaking locations in the non-sparse case, is given by:
\begin{align*}
\notag
P(\mu_{k} | \cdots) & \propto \mathcal{N}(\mu_{k} | c,d) \cdot \prod_{(j,t) : \gamma_{j,t} = 1, d_{j,t} =k} \exp\left\lbrace -\frac{1}{2 \lambda_{j}} (\beta_{j,t} -\mu_{k})^2 \right\rbrace \\ \notag
 & \propto \exp\left\lbrace -\frac{1}{2d} (\mu_{k}^2 -2c\mu_{k}) \right\rbrace \exp\left\lbrace -\frac{1}{2} \sum_{(j,t) : \gamma_{j,t} = 1, d_{j,t} =k} \mu_{k}^2\lambda_{j}^{-1} -2\mu_{k}\lambda_{j}^{-1} \beta_{j,t} \right\rbrace \\ \notag
 & \propto \exp\left\lbrace -\frac{1}{2} \left[ \mu_{k}^2 d^{-1} -2c\mu_{k} d^{-1} + \sum_{(j,t) : \gamma_{j,t} = 1, d_{j,t} =k} \mu_{k}^2 \lambda_{j}^{-1} -2\mu_{k} \lambda_{j}^{-1} \beta_{j,t} \right] \right\rbrace \\
 & \propto \mathcal{N}(\overline{c}, \overline{d}),
\end{align*}
where
\begin{equation*}
\overline{d} = \left( d^{-1} + \sum_{(j,t) : \gamma_{j,t} = 1,d_{j,t} =k} \lambda_{j}^{-1} \right)^{-1}, \qquad \overline{c} = \overline{d} \left( d^{-1} c + \sum_{(j,t) : \gamma_{j,t} = 1, d_{j,t} =k} \lambda_{j}^{-1} \beta_{j,t} \right).
\end{equation*}
and for the scale parameter we have the following representation:
\begin{align*}
\notag
P(\tau_{k} | \cdots) & \propto \mathcal{G}a(\tau_{k} | a_1,b_1) \cdot \prod_{(j,t) : \gamma_{j,t}=1, d_{j,t} =k} \frac{\tau_{k}}{2} \exp\left\lbrace -\frac{\tau_{k}}{2} \lambda_{j} \right\rbrace \\ \notag
 & \propto \tau_{k}^{a_1-1} \exp\left\lbrace -\frac{\tau_{k}}{b_1} \right\rbrace \cdot \prod_{(j,t) : \gamma_{j,t}=1,, d_{j,t} =k} \frac{\tau_{k}}{2} \exp\left\lbrace -\frac{\tau_{k}}{2} \lambda_{j} \right\rbrace \\ \notag
 & \propto \tau_{k}^{a_1-1} \exp\left\lbrace -\frac{\tau_{k}}{b_1} \right\rbrace \tau_{k}^{m_1} \exp\left\lbrace -\frac{\tau_{k}}{2} \sum_{(j,t) : \gamma_{j,t}=1, d_{j,t} =k} \lambda_{j} \right\rbrace \\ \notag
 & \propto \tau_{k}^{a_1+m_1-1} \exp\left\lbrace -\tau_{k} \left( \frac{1}{b_1} + \frac{1}{2} \sum_{(j,t) : \gamma_{j,t}=1, d_{j,t} =k} \lambda_{j} \right) \right\rbrace \\
 & \propto \mathcal{G}a(\overline{a}_1, \overline{b}_1),
\end{align*}
with
\begin{equation*}
\overline{a}_1 = a_1 + m_1, \qquad \overline{b}_1 = \frac{2b_1}{2+b_1 \sum_{(j,t) : \gamma_{j,t}=1, d_{j,t} =k} \lambda_{j}},
\end{equation*}
where $m_1 = \sum_{(j,t)} \mathbb{I}(\gamma_{j,t}=1) \mathbb{I}(d_{j,t} = k) = \sum_{j=1}^{n^2} \sum_{t=1}^T \gamma_{j,t}\mathbb{I}(d_{j,t} = k)$.

\subsection{Posterior of the allocation variables $(d_{j,t},\gamma_{j,t})$}
Regarding the allocation variables $d_{j,t},\gamma_{j,t}$, we obtain the following full conditional for the sparse and non-sparse case. Let $u_{j,t} \sim \mathcal{U}([0,1])$ for all $j=1,\ldots,n^2$ and all $t=1,\ldots,T$, be the auxiliary slice sampling variable. In the non-sparse case, we have:
\begin{align*}
\notag
P(d_{j,t} = k, \gamma_{j,t} = 1 |\cdots) & \propto (1-\pi_{t}) \mathcal{N}(\beta_{j,t}| \mu_k,\lambda_{j}) \mathcal{E}xp \left( \lambda_{j}| \frac{2}{\tau_{k}} \right) \mathbb{I}(u_{j,t} < w_k) \\
 & \propto \frac{(1-\pi_t) \mathcal{N}(\beta_{j,t}| \mu_{k},\lambda_{j}) \mathcal{E}xp\left( \lambda_{j}| \frac{2}{\tau_{k}} \right)}{\sum_{i\in A_w(u_{j,t})} \mathcal{N}(\beta_{j,t}| \mu_{i},\lambda_{j}) \mathcal{E}xp\left( \lambda_{j} | \frac{2}{\tau_{i}} \right)}, \quad \forall \, k \in A_w(u_{j,t}),
\end{align*}
where $A_w(u_{j,t}) = \lbrace k \in \lbrace 1,2,\ldots,k^* \rbrace : w_k > u_{j,t} \rbrace$, where $k^*$ is the number of ties.
On the other hand, in the sparse case we have the non-normalised posterior probability
\begin{equation*}
P(d_{j,t} = 0, \gamma_{j,t} = 0 |\cdots) \propto \pi_t \mathcal{N}(\beta_{j,t}|0,\lambda_{j}) \mathcal{E}xp\left( \lambda_{j}|\frac{2}{\tau_{0}} \right) \mathbb{I}(u_{j,t} < 1),
\end{equation*}
and full conditional distribution has the following representation:
\begin{equation*}
P(d_{j,t} = k, \gamma_{j,t} = 0 |\cdots) \propto \left\lbrace \begin{array}{ll}
\pi_t \mathcal{N}(\beta_{j,t}|0,\lambda_{j}) \mathcal{E}xp\left( \lambda_{j}|\frac{2}{\tau_{0}} \right) & \mathrm{if} \; k = 0, \\
0 & \mathrm{if} \; k > 0.
\end{array} \right.
\end{equation*}

\subsection{Posterior for $\beta_{j,t}$}
Posterior distribution for $\boldsymbol{\beta}_t$, for all $t=2,\ldots,T$. Let $\boldsymbol{\mu}_t^* = (\mu_{d_{1,t}},\dots,\mu_{d_{n^2,t}})'$ and $\boldsymbol{\gamma}_t = (\gamma_{1,t},\dots,\gamma_{n^2,t})'$, and denote the Hadamard product by $\odot$. We have:
\begin{align*}
\notag
P(\boldsymbol{\beta}_t|\cdots) & \propto (2\pi)^{-n^2/2} \abs{\boldsymbol{\Lambda}}^{-1/2} \exp\left\lbrace -\frac{1}{2} (\boldsymbol{\beta}_t-\boldsymbol{\mu}_t)' \boldsymbol{\Lambda}^{-1} (\boldsymbol{\beta}_t-\boldsymbol{\mu}_t) \right\rbrace \\ \notag
 & \quad \cdot (2\pi)^{-n^2/2} \abs{\boldsymbol{\Sigma}}^{-1/2} \exp\left\lbrace -\frac{1}{2} (\mathbf{y}_t -\mathbf{X}_t \boldsymbol{\beta}_t)' \boldsymbol{\Sigma}_{\epsilon}^{-1} (\mathbf{y}_t -\mathbf{X}_t \boldsymbol{\beta}_t) \right\rbrace \\ \notag
 & \propto \exp\left\lbrace -\frac{1}{2} (\boldsymbol{\beta}_t' \boldsymbol{\Lambda}^{-1}\boldsymbol{\beta}_t -2\boldsymbol{\beta}_t' \boldsymbol{\Lambda}^{-1} \boldsymbol{\mu}_t) \right\rbrace \cdot \exp\left\lbrace -\frac{1}{2} \left( \boldsymbol{\beta}_t' \mathbf{X}_t' \boldsymbol{\Sigma}_{\epsilon}^{-1} \mathbf{X}_t \boldsymbol{\beta}_t -2\boldsymbol{\beta}_t' \mathbf{X}_t' \boldsymbol{\Sigma}_{\epsilon}^{-1} \mathbf{y}_t \right) \right\rbrace \\ \notag
 & \propto \exp\left\lbrace -\frac{1}{2} \left[ \sum_{j=1}^{n^2} \frac{1}{\lambda_{j}} \left( \beta_{j,t}^2 - 2\beta_{j,t} \mu_{d_{j,t}} \gamma_{j,t} \right) +\left( \boldsymbol{\beta}_t' \mathbf{X}_t' \boldsymbol{\Sigma}_{\epsilon}^{-1} \mathbf{X}_t \boldsymbol{\beta}_t -2\boldsymbol{\beta}_t' \mathbf{X}_t' \boldsymbol{\Sigma}_{\epsilon}^{-1} \mathbf{y}_t \right) \right] \right\rbrace \\ \notag
 & \propto \exp\left\lbrace -\frac{1}{2} \left[ \left( \boldsymbol{\beta}_t' \boldsymbol{\Lambda}^{-1} \boldsymbol{\beta}_t -2\boldsymbol{\beta}_t' \boldsymbol{\Lambda}^{-1} (\boldsymbol{\mu}_t^* \odot \boldsymbol{\gamma}_t) \right) + \boldsymbol{\beta}_t' \mathbf{X}_t' \boldsymbol{\Sigma}^{-1} \mathbf{X}_t \boldsymbol{\beta}_t -2\boldsymbol{\beta}_t' \mathbf{X}_t' \boldsymbol{\Sigma}^{-1} \mathbf{y}_t \right] \right\rbrace \\
 & \propto \mathcal{N}(\overline{\boldsymbol{\mu}}_t,\overline{\Sigma}_t),
\end{align*}
where
\begin{equation*}
\overline{\Sigma}_t = \left( \boldsymbol{\Lambda}^{-1} + \mathbf{X}_t' \boldsymbol{\Sigma}^{-1} \mathbf{X}_t \right)^{-1}, \qquad \overline{\boldsymbol{\mu}}_t = \overline{\Sigma}_t \left( \boldsymbol{\Lambda}^{-1} (\boldsymbol{\mu}_t^* \odot \boldsymbol{\gamma}_t) + \mathbf{X}_t' \boldsymbol{\Sigma}^{-1} \mathbf{y}_t \right).
\end{equation*}

\subsection{Posterior for covariance matrix $\Sigma$}
Posterior distribution for the covariance matrix $\boldsymbol{\Sigma}$.
\begin{align*}
\notag
P(\boldsymbol{\Sigma} | \cdots) & \propto \mathcal{IW}(\nu,\boldsymbol{\Psi}) \cdot \prod_{t=1}^T \abs{\boldsymbol{\Sigma}}^{-1/2} \exp\left\lbrace -\frac{1}{2} (\mathbf{y}_t - \mathbf{X}_t \boldsymbol{\beta}_t)' \boldsymbol{\Sigma}^{-1} (\mathbf{y}_t - \mathbf{X}_t \boldsymbol{\beta}_t) \right\rbrace \\ \notag
 & \propto \abs{\boldsymbol{\Sigma}}^{-(\nu+n+1)/2} \exp\left\lbrace -\frac{1}{2} \operatorname{tr}\left( \boldsymbol{\Psi} \boldsymbol{\Sigma}^{-1} \right) \right\rbrace \abs{\boldsymbol{\Sigma}}^{-T/2} \exp\left\lbrace -\frac{1}{2} \sum_{t=1}^T \operatorname{tr}\left( (\mathbf{y}_t - \mathbf{X}_t \boldsymbol{\beta}_t)' \boldsymbol{\Sigma}^{-1} (\mathbf{y}_t - \mathbf{X}_t \boldsymbol{\beta}_t) \right) \right\rbrace \\
 & \propto \mathcal{IW}(\overline{\nu}, \overline{\boldsymbol{\Psi}}),
\end{align*}
where
\begin{equation*}
\overline{\nu} = \nu+T, \qquad \overline{\boldsymbol{\Psi}} = \boldsymbol{\Psi} + \sum_{t=1}^T (\mathbf{y}_t - \mathbf{X}_t \boldsymbol{\beta}_t)(\mathbf{y}_t - \mathbf{X}_t \boldsymbol{\beta}_t)'.
\end{equation*}

\subsection{Posterior for mixing probability $\pi_t$}
Posterior distribution for the mixing probability $\pi_t$, for all $t=1,\ldots,T$.
\begin{align*}
\notag
P(\pi_t | \cdots) & \propto \mathcal{B}e(1,\eta) \cdot \prod_{j=1}^{n^2} \pi_t^{\mathbb{I}(\gamma_{j,t}=0)} (1-\pi_t)^{\mathbb{I}(\gamma_{j,t}=1)} \\ \notag
 & \propto (1-\pi_t)^{\eta-1} \: \pi_t^{n^2-\sum_{j=1}^{n^2} \mathbb{I}(\gamma_{j,t}=1)} \: (1-\pi_t)^{\sum_{j=1}^{n^2} \mathbb{I}(\gamma_{j,t}=1)} \\
 & \propto \mathcal{B}e(\overline{\phi}, \overline{\eta}),
\end{align*}
where
\begin{equation*}
\overline{\phi} = 1+n^2 -\sum_{j=1}^{n^2} \mathbb{I}(\gamma_{j,t}=1), \qquad \overline{\eta} = \eta +\sum_{j=1}^{n^2} \mathbb{I}(\gamma_{j,t}=1).
\end{equation*}

\clearpage

\section{Posterior distributions: diffuse Normal spike}
\subsection{Posterior for stick breaking un-normalised weights $v_{i,t}$}
Posterior distribution for $v_{i,t}$, for all $t=1,\ldots,T$ and all $k=1,\ldots,k^*$, where $k^*$ is the number of ties. We use the convention $z_{i,0} = m_{i,0} = 0$, for all $i=1,2,\ldots$.
\begin{align*}
\notag
P(v_{k,t} | z_{k,t},z_{k,t-1}, \ldots) & \propto P(v_{k,t}|z_{k,t-1}) P(z_{k,t}|v_{k,t}) P(\ldots | v_{k,t}) \\ \notag
 & \propto v_{k,t}^{z_{k,t-1}} (1-v_{k,t})^{\alpha+m_{k,t-1}-z_{k,t-1}-1} \frac{m_{k,t}!}{z_{k,t}! (m_{k,t}-z_{k,t})!} v_{k,t}^{z_{k,t}} (1-v_{k,t})^{m_{k,t}-z_{k,t}} \\ \notag
 & \quad \cdot \left[ \prod_{j : d_{j,t} = k, \: \gamma_{j,t}=1} v_{j,t} \right] \left[ \prod_{j : d_{j,t} >k, \: \gamma_{j,t}=1} (1-v_{j,t}) \right] \\
 & \propto \mathcal{B}e(\overline{\xi}_t, \overline{\alpha}_t),
\end{align*}
where
\begin{align*}
\overline{\xi}_t & = 1+z_{k,t}+z_{k,t-1} + \sum_{j} \mathbb{I}(d_{j,t} = k, \gamma_{j,t} = 1), \\
\overline{\alpha}_t & = \alpha +m_{k,t}-z_{k,t} +m_{k,t-1}-z_{k,t-1} + \sum_{j} \mathbb{I}(d_{j,t} > k, \gamma_{j,t} = 1).
\end{align*}
The normalised weights are then computed by $w_{k,t} = v_{k,t} \prod_{j < k} (1-v_{j,t})$.
The posterior distribution of the latent auxiliary variables $z_{k,t}$, for all $t=1,\ldots,T-1$ and all $k=1,\ldots,k^*$ is given by
\begin{align*}
\notag
P(z_{k,t} | v_{k,t+1},v_{k,t}) & \propto P(z_{k,t} | v_{k,t}) P(v_{k,t+1} | z_{k,t}) \\ \notag
 & \propto \frac{m_{k,t}!}{z_{k,t}! (m_{k,t}-z_{k,t})!} v_{k,t}^{z_{k,t}} (1-v_{k,t})^{m_{k,t}-z_{k,t}} \\ \notag
 & \quad \cdot \frac{\Gamma(\alpha+m_{k,t}+1)}{\Gamma(z_{k,t}+1)\Gamma(\alpha+m_{k,t}-z_{k,t})} v_{k,t+1}^{z_{k,t}} (1-v_{k,t+1})^{\alpha+m_{k,t}-z_{k,t}-1} \\
 & \propto \frac{1}{(\Gamma(z_{k,t}+1))^2 \Gamma(\alpha+m_{k,t}-z_{k,t}) \Gamma(m_{k,t}-z_{k,t}+1)} \left(\frac{v_{k,t} v_{k,t+1}}{(1-v_{k,t})(1-v_{k,t+1})} \right)^{z_{k,t}}.
\end{align*}
Finally, the auxiliary variable for the slice sampler has posterior distribution given by
\begin{equation*}
P(u_{j,t}| \cdots) \propto \left\{
\begin{array}{ll}
\mathbb{I}(u_{j,t} < w_{d_{j,t}})^{\gamma_{j,t}}  	& \mathrm{if} \; \gamma_{j,t}=1 ,\\
\mathbb{I}(u_{j,t} < 1)^{1-\gamma_{j,t}} 			& \mathrm{if} \; \gamma_{j,t}=0.
\end{array} \right.
\end{equation*}

\subsection{Posterior for $\lambda_{j}$}
Posterior distribution for $\lambda_{j}$, for $j=1,\ldots,n^2$. Define $\mu_{k_j}^*,\tau_{k_j}^*$ are the location and scale of $\beta_{j,t}$, respectively, when sparse or non-sparse component is chosen.
\begin{align*}
\notag
P(\lambda_{j}|\cdots) & \propto \prod_{t=2}^T \lambda_{j}^{-\gamma_{j,t}/2} \exp\left\lbrace -\gamma_{j,t} \frac{1}{2\lambda_{j}} (\beta_{j,t} -\mu_{d_{j,t}})^2 \right\rbrace \exp\left\lbrace \gamma_{j,t} \left( -\frac{\lambda_{j} \tau_{d_{j,t}}}{2} \right) \right\rbrace \\ \notag
 & \propto \lambda_{j}^{-\frac{1}{2}\sum_{t=2}^T \gamma_{j,t}} \exp\left\lbrace -\frac{1}{2}  \left[ \lambda_{j} \left( \sum_{t=2}^T \gamma_{j,t} \tau_{d_{j,t}} \right) + \frac{1}{\lambda_j} \left( \sum_{t=2}^T \gamma_{j,t} (\beta_{j,t}-\mu_{d_{j,t}})^2 \right) \right] \right\rbrace \\
 & \propto GiG(p,a,b)
\end{align*}
where
\begin{equation*}
p= 1-\sum_{t=2}^T \gamma_{j,t}, \qquad a= \sum_{t=2}^T \gamma_{j,t} \tau_{d_{j,t}}, \qquad b= \sum_{t=2}^T \gamma_{j,t} (\beta_{j,t}-\mu_{d_{j,t}})^2.
\end{equation*}
The probability density function of a generalised inverse Gaussian, for $p\in \mathbb{R}, a >0, b >0$ and $K_p(\cdot)$ being a modified Bessel function of the second kind, is
\begin{equation*}
GiG(x|p,a,b) = \frac{(a/b)^{p/2}}{2 K_p(\sqrt{ab})} x^{p-1} e^{-(ax+b/x)/2}, \qquad x \in (0,\infty).
\end{equation*}
The vector $\boldsymbol{\lambda} = (\lambda_{1},\ldots,\lambda_{n^2})$ represents the diagonal of the diagonal matrix $\boldsymbol{\Lambda}$ of size $(n^2 \times n^2)$.

\subsection{Posterior of the stick breaking locations $(\mu_{k},\tau_{k})$}
The posterior distribution for the stick breaking locations in the sparse case are given by:
\begin{equation*}
\mu_{0} | \cdots \sim \delta_{(0)}(\mu_{0}).
\end{equation*}
and
\begin{align*}
\notag
P(\tau_{0} | \cdots) & \propto \mathcal{IG}(\tau_{0} | a_0,b_0) \cdot \prod_{(j,t) : \gamma_{j,t}=0} \tau_{0}^{-1/2} \exp\left\lbrace -\frac{\beta_{j,t}^2}{2\tau_0} \right\rbrace \\ \notag
 & \propto \tau_{0}^{-a_0-1} \exp\left\lbrace -\frac{b_0}{\tau_0} \right\rbrace \tau_{0}^{-m_0/2} \exp\left\lbrace -\frac{1}{2\tau_{0}} \sum_{(j,t) : \gamma_{j,t}=0} \beta_{j,t}^2 \right\rbrace \\ \notag
 & \propto \tau_{0}^{-a_0-m_0/2-1} \exp\left\lbrace -\frac{1}{\tau_{0}} \left( b_0 + \frac{1}{2} \sum_{(j,t) : \gamma_{j,t}=0} \beta_{j,t}^2 \right) \right\rbrace \\
 & \propto \mathcal{IG}(\overline{a}_0, \overline{b}_0),
\end{align*}
with
\begin{equation*}
\overline{a}_0 = a_0 + m_0/2, \qquad \overline{b}_0 = b_0 + \frac{1}{2} \sum_{(j,t) : \gamma_{j,t}=0} \beta_{j,t}^2,
\end{equation*}
where $m_0 = \sum_{(j,t)} \mathbb{I}(\gamma_{j,t}=0) = \sum_{j=1}^{n^2} \sum_{t=1}^T (1-\gamma_{j,t})$ and we used the parametrisation of the Inverse Gamma distribution with shape $a >0$ and scale $b >0$, that is
\begin{equation*}
\mathcal{IG}(x|a,b) = \frac{b^a}{\Gamma(a)} x^{-a-1} e^{-b/x}, \qquad x \in (0,\infty).
\end{equation*}

Regarding the sample in the non-sparse case, we generate $(\mu_k, \tau_k)$ for $k = 1,\ldots, N^{\ast}$ and we
have the following full conditional separately for $\mu_k$ and $\tau_k$.  The posterior distribution for the stick breaking locations in the non-sparse case, is given by:
\begin{align*}
\notag
P(\mu_{k} | \cdots) & \propto \mathcal{N}(\mu_{k} | c,d) \cdot \prod_{(j,t) : \gamma_{j,t} = 1, d_{j,t} =k} \exp\left\lbrace -\frac{1}{2 \lambda_{j}} (\beta_{j,t} -\mu_{k})^2 \right\rbrace \\ \notag
 & \propto \exp\left\lbrace -\frac{1}{2d} (\mu_{k}^2 -2c\mu_{k}) \right\rbrace \exp\left\lbrace -\frac{1}{2} \sum_{(j,t) : \gamma_{j,t} = 1, d_{j,t} =k} \mu_{k}^2\lambda_{j}^{-1} -2\mu_{k}\lambda_{j}^{-1} \beta_{j,t} \right\rbrace \\ \notag
 & \propto \exp\left\lbrace -\frac{1}{2} \left[ \mu_{k}^2 d^{-1} -2c\mu_{k} d^{-1} + \sum_{(j,t) : \gamma_{j,t} = 1, d_{j,t} =k} \mu_{k}^2 \lambda_{j}^{-1} -2\mu_{k} \lambda_{j}^{-1} \beta_{j,t} \right] \right\rbrace \\
 & \propto \mathcal{N}(\overline{c}, \overline{d}),
\end{align*}
where
\begin{equation*}
\overline{d} = \left( d^{-1} + \sum_{(j,t) : \gamma_{j,t} = 1,d_{j,t} =k} \lambda_{j}^{-1} \right)^{-1}, \qquad \overline{c} = \overline{d} \left( d^{-1} c + \sum_{(j,t) : \gamma_{j,t} = 1, d_{j,t} =k} \lambda_{j}^{-1} \beta_{j,t} \right).
\end{equation*}
and for the scale parameter we have the following representation:
\begin{align*}
\notag
P(\tau_{k} | \cdots) & \propto \mathcal{G}a(\tau_{k} | a_1,b_1) \cdot \prod_{(j,t) : \gamma_{j,t}=1, d_{j,t} =k} \frac{\tau_{k}}{2} \exp\left\lbrace -\frac{\tau_{k}}{2} \lambda_{j} \right\rbrace \\ \notag
 & \propto \tau_{k}^{a_1-1} \exp\left\lbrace -\frac{\tau_{k}}{b_1} \right\rbrace \cdot \prod_{(j,t) : \gamma_{j,t}=1,, d_{j,t} =k} \frac{\tau_{k}}{2} \exp\left\lbrace -\frac{\tau_{k}}{2} \lambda_{j} \right\rbrace \\ \notag
 & \propto \tau_{k}^{a_1-1} \exp\left\lbrace -\frac{\tau_{k}}{b_1} \right\rbrace \tau_{k}^{m_1} \exp\left\lbrace -\frac{\tau_{k}}{2} \sum_{(j,t) : \gamma_{j,t}=1, d_{j,t} =k} \lambda_{j} \right\rbrace \\ \notag
 & \propto \tau_{k}^{a_1+m_1-1} \exp\left\lbrace -\tau_{k} \left( \frac{1}{b_1} + \frac{1}{2} \sum_{(j,t) : \gamma_{j,t}=1, d_{j,t} =k} \lambda_{j} \right) \right\rbrace \\
 & \propto \mathcal{G}a(\overline{a}_1, \overline{b}_1),
\end{align*}
with
\begin{equation*}
\overline{a}_1 = a_1 + m_1, \qquad \overline{b}_1 = \frac{2b_1}{2+b_1 \sum_{(j,t) : \gamma_{j,t}=1, d_{j,t} =k} \lambda_{j}},
\end{equation*}
where $m_1 = \sum_{(j,t)} \mathbb{I}(\gamma_{j,t}=1) \mathbb{I}(d_{j,t} = k) = \sum_{j=1}^{n^2} \sum_{t=1}^T \gamma_{j,t}\mathbb{I}(d_{j,t} = k)$.

\subsection{Posterior of the allocation variables $(d_{j,t},\gamma_{j,t})$}
Regarding the allocation variables $d_{j,t},\gamma_{j,t}$, we obtain the following full conditional for the sparse and non-sparse case. Let $u_{j,t} \sim \mathcal{U}([0,1])$ for all $j=1,\ldots,n^2$ and all $t=1,\ldots,T$, be the auxiliary slice sampling variable. In the non-sparse case, we have:
\begin{align*}
\notag
P(d_{j,t} = k, \gamma_{j,t} = 1 |\cdots) & \propto (1-\pi_{t}) \mathcal{N}(\beta_{j,t}| \mu_k,\lambda_{j}) \mathcal{E}xp \left( \lambda_{j}| \frac{2}{\tau_{k}} \right) \mathbb{I}(u_{j,t} < w_k) \\
 & \propto \frac{(1-\pi_t) \mathcal{N}(\beta_{j,t}| \mu_{k},\lambda_{j}) \mathcal{E}xp\left( \lambda_{j}| \frac{2}{\tau_{k}} \right)}{\sum_{i\in A_w(u_{j,t})} \mathcal{N}(\beta_{j,t}| \mu_{i},\lambda_{j}) \mathcal{E}xp\left( \lambda_{j} | \frac{2}{\tau_{i}} \right)}, \quad \forall \, k \in A_w(u_{j,t}),
\end{align*}
where $A_w(u_{j,t}) = \lbrace k \in \lbrace 1,2,\ldots,k^* \rbrace : w_k > u_{j,t} \rbrace$, where $k^*$ is the number of ties.
On the other hand, in the sparse case we have the non-normalised posterior probability
\begin{equation*}
P(d_{j,t} = 0, \gamma_{j,t} = 0 |\cdots) \propto \pi_t \mathcal{N}(\beta_{j,t}|0,\tau_0) \mathbb{I}(u_{j,t} < 1),
\end{equation*}
and full conditional distribution has the following representation:
\begin{equation*}
P(d_{j,t} = k, \gamma_{j,t} = 0 |\cdots) \propto \left\lbrace \begin{array}{ll}
\pi_t \mathcal{N}(\beta_{j,t}|0,\tau_0) & \mathrm{if} \; k = 0, \\
0 & \mathrm{if} \; k > 0.
\end{array} \right.
\end{equation*}

\subsection{Posterior for $\beta_{j,t}$}
Posterior distribution for $\boldsymbol{\beta}_t$, for all $t=2,\ldots,T$. Let $\boldsymbol{\mu}_t^* = (\mu_{d_{1,t}},\dots,\mu_{d_{n^2,t}})'$ and $\boldsymbol{\gamma}_t = (\gamma_{1,t},\dots,\gamma_{n^2,t})'$, and denote the Hadamard product by $\odot$. We have:
\begin{align*}
\notag
P(\boldsymbol{\beta}_t&|\cdots)  \propto (2\pi)^{-n^2/2} \abs{\boldsymbol{\Lambda}}^{-1/2} \exp\left\lbrace -\frac{1}{2} (\boldsymbol{\beta}_t-\boldsymbol{\mu}_t)' \boldsymbol{\Lambda}^{-1} (\boldsymbol{\beta}_t-\boldsymbol{\mu}_t) \right\rbrace \\ \notag
 & \quad \cdot (2\pi)^{-n^2/2} \abs{\boldsymbol{\Sigma}}^{-1/2} \exp\left\lbrace -\frac{1}{2} (\mathbf{y}_t -\mathbf{X}_t \boldsymbol{\beta}_t)' \boldsymbol{\Sigma}_{\epsilon}^{-1} (\mathbf{y}_t -\mathbf{X}_t \boldsymbol{\beta}_t) \right\rbrace \\ \notag
% & \propto \exp\left\lbrace -\frac{1}{2} (\boldsymbol{\beta}_t' \boldsymbol{\Lambda}^{-1}\boldsymbol{\beta}_t -2\boldsymbol{\beta}_t' \boldsymbol{\Lambda}^{-1} \boldsymbol{\mu}_t) \right\rbrace \cdot \exp\left\lbrace -\frac{1}{2} \left( \boldsymbol{\beta}_t' \mathbf{X}_t' \boldsymbol{\Sigma}_{\epsilon}^{-1} \mathbf{X}_t \boldsymbol{\beta}_t -2\boldsymbol{\beta}_t' \mathbf{X}_t' \boldsymbol{\Sigma}_{\epsilon}^{-1} \mathbf{y}_t \right) \right\rbrace \\ \notag
 & \propto \exp\left\lbrace -\frac{1}{2} \left[ \sum_{j=1}^{n^2} \left( \frac{\beta_{j,t}^2(1-\gamma_{j,t})}{\tau_0} + \frac{\beta_{j,t}^2 \gamma_{j,t}}{\lambda_j} - 2\frac{\beta_{j,t} \mu_{d_{j,t}} \gamma_{j,t}}{\lambda_{j}} \right) +\left( \boldsymbol{\beta}_t' \mathbf{X}_t' \boldsymbol{\Sigma}_{\epsilon}^{-1} \mathbf{X}_t \boldsymbol{\beta}_t -2\boldsymbol{\beta}_t' \mathbf{X}_t' \boldsymbol{\Sigma}_{\epsilon}^{-1} \mathbf{y}_t \right) \right] \right\rbrace \\ \notag
 & \propto \exp\Biggl\lbrace -\frac{1}{2} \Biggl[ \left( (\boldsymbol{\beta}_t \odot (1-\boldsymbol{\gamma}_t))' \tau_0^{-1} (\boldsymbol{\beta}_t \odot (1-\boldsymbol{\gamma}_t)) + (\boldsymbol{\beta}_t \odot \boldsymbol{\gamma}_t)' \boldsymbol{\Lambda}^{-1} (\boldsymbol{\beta}_t \odot \boldsymbol{\gamma}_t) -2\boldsymbol{\beta}_t' \boldsymbol{\Lambda}^{-1} (\boldsymbol{\mu}_t^* \odot \boldsymbol{\gamma}_t) \right) \\ 
 & \quad + \boldsymbol{\beta}_t' \mathbf{X}_t' \boldsymbol{\Sigma}^{-1} \mathbf{X}_t \boldsymbol{\beta}_t -2\boldsymbol{\beta}_t' \mathbf{X}_t' \boldsymbol{\Sigma}^{-1} \mathbf{y}_t \Biggr] \Biggr\rbrace \\
 & \propto \mathcal{N}(\overline{\boldsymbol{\mu}}_t,\overline{\Sigma}_t),
\end{align*}
where
\begin{equation*}
\overline{\Sigma}_t = \left( \boldsymbol{\Lambda}_{\tau_0,\gamma}^{-1} + \mathbf{X}_t' \boldsymbol{\Sigma}^{-1} \mathbf{X}_t \right)^{-1}, \qquad \overline{\boldsymbol{\mu}}_t = \overline{\Sigma}_t \left( \boldsymbol{\Lambda}^{-1} (\boldsymbol{\mu}_t^* \odot \boldsymbol{\gamma}_t) + \mathbf{X}_t' \boldsymbol{\Sigma}^{-1} \mathbf{y}_t \right)
\end{equation*}
and $\boldsymbol{\Lambda}_{\tau_0,\gamma}$ is a $n^2 \times n^2$ diagonal matrix whose $jj$-th entry is
\begin{equation*}
\boldsymbol{\Lambda}_{\tau_0,\gamma,jj} = \begin{cases}
\lambda_j & \text{if } \gamma_{j,t} = 1, \\
\tau_0 & \text{if } \gamma_{j,t} = 0.
\end{cases}
\end{equation*}

\subsection{Posterior for covariance matrix $\Sigma$}
Posterior distribution for the covariance matrix $\boldsymbol{\Sigma}$.
\begin{align*}
\notag
P(\boldsymbol{\Sigma} | \cdots) & \propto \mathcal{IW}(\nu,\boldsymbol{\Psi}) \cdot \prod_{t=1}^T \abs{\boldsymbol{\Sigma}}^{-1/2} \exp\left\lbrace -\frac{1}{2} (\mathbf{y}_t - \mathbf{X}_t \boldsymbol{\beta}_t)' \boldsymbol{\Sigma}^{-1} (\mathbf{y}_t - \mathbf{X}_t \boldsymbol{\beta}_t) \right\rbrace \\ \notag
 & \propto \abs{\boldsymbol{\Sigma}}^{-(\nu+n+1)/2} \exp\left\lbrace -\frac{1}{2} \operatorname{tr}\left( \boldsymbol{\Psi} \boldsymbol{\Sigma}^{-1} \right) \right\rbrace \abs{\boldsymbol{\Sigma}}^{-T/2} \exp\left\lbrace -\frac{1}{2} \sum_{t=1}^T \operatorname{tr}\left( (\mathbf{y}_t - \mathbf{X}_t \boldsymbol{\beta}_t)' \boldsymbol{\Sigma}^{-1} (\mathbf{y}_t - \mathbf{X}_t \boldsymbol{\beta}_t) \right) \right\rbrace \\
 & \propto \mathcal{IW}(\overline{\nu}, \overline{\boldsymbol{\Psi}}),
\end{align*}
where
\begin{equation*}
\overline{\nu} = \nu+T, \qquad \overline{\boldsymbol{\Psi}} = \boldsymbol{\Psi} + \sum_{t=1}^T (\mathbf{y}_t - \mathbf{X}_t \boldsymbol{\beta}_t)(\mathbf{y}_t - \mathbf{X}_t \boldsymbol{\beta}_t)'.
\end{equation*}

\subsection{Posterior for mixing probability $\pi_t$}
Posterior distribution for the mixing probability $\pi_t$, for all $t=1,\ldots,T$.
\begin{align*}
\notag
P(\pi_t | \cdots) & \propto \mathcal{B}e(1,\eta) \cdot \prod_{j=1}^{n^2} \pi_t^{\mathbb{I}(\gamma_{j,t}=0)} (1-\pi_t)^{\mathbb{I}(\gamma_{j,t}=1)} \\ \notag
 & \propto (1-\pi_t)^{\eta-1} \: \pi_t^{n^2-\sum_{j=1}^{n^2} \mathbb{I}(\gamma_{j,t}=1)} \: (1-\pi_t)^{\sum_{j=1}^{n^2} \mathbb{I}(\gamma_{j,t}=1)} \\
 & \propto \mathcal{B}e(\overline{\phi}, \overline{\eta}),
\end{align*}
where
\begin{equation*}
\overline{\phi} = 1+n^2 -\sum_{j=1}^{n^2} \mathbb{I}(\gamma_{j,t}=1), \qquad \overline{\eta} = \eta +\sum_{j=1}^{n^2} \mathbb{I}(\gamma_{j,t}=1).
\end{equation*}

\clearpage

\section{Posterior distributions: Dirac spike} \label{sec:apdx_posterior_Dirac}
\subsection{Posterior for stick breaking un-normalised weights $v_{i,t}$}
Posterior distribution for $v_{i,t}$, for all $t=2,\ldots,T$ and all $k=1,\ldots,k^*$, where $k^*$ is the number of ties. We use the convention $z_{i,0} = m_{i,0} = 0$, for all $i=1,2,\ldots$.
\begin{align*}
\notag
P(v_{k,t} | z_{k,t},z_{k,t-1}, \ldots) & \propto P(v_{k,t}|z_{k,t-1}) P(z_{k,t}|v_{k,t}) P(\ldots | v_{k,t}) \\ \notag
 & \propto v_{k,t}^{z_{k,t-1}} (1-v_{k,t})^{\alpha+m_{k,t-1}-z_{k,t-1}-1} \frac{m_{k,t}!}{z_{k,t}! (m_{k,t}-z_{k,t})!} v_{k,t}^{z_{k,t}} (1-v_{k,t})^{m_{k,t}-z_{k,t}} \\ \notag
 & \quad \cdot \left[ \prod_{j : d_{j,t} = k, \: \gamma_{j,t}=1} v_{j,t} \right] \left[ \prod_{j : d_{j,t} >k, \: \gamma_{j,t}=1} (1-v_{j,t}) \right] \\
 & \propto \mathcal{B}e(\overline{\xi}_t, \overline{\alpha}_t),
\end{align*}
where
\begin{align*}
\overline{\xi}_t & = 1+z_{k,t}+z_{k,t-1} + \sum_{j} \mathbb{I}(d_{j,t} = k, \gamma_{j,t} = 1), \\
\overline{\alpha}_t & = \alpha +m_{k,t}-z_{k,t} +m_{k,t-1}-z_{k,t-1} + \sum_{j} \mathbb{I}(d_{j,t} > k, \gamma_{j,t} = 1).
\end{align*}
The normalised weights are then computed by $w_{k,t} = v_{k,t} \prod_{j < k} (1-v_{j,t})$.
The posterior distribution of the latent auxiliary variables $z_{k,t}$, for all $t=1,\ldots,T-1$ and all $k=1,\ldots,k^*$ is given by
\begin{align*}
\notag
P(z_{k,t} | v_{k,t+1},v_{k,t}) & \propto P(z_{k,t} | v_{k,t}) P(v_{k,t+1} | z_{k,t}) \\ \notag
 & \propto \frac{m_{k,t}!}{z_{k,t}! (m_{k,t}-z_{k,t})!} v_{k,t}^{z_{k,t}} (1-v_{k,t})^{m_{k,t}-z_{k,t}} \\ \notag
 & \quad \cdot \frac{\Gamma(\alpha+m_{k,t}+1)}{\Gamma(z_{k,t}+1)\Gamma(\alpha+m_{k,t}-z_{k,t})} v_{k,t+1}^{z_{k,t}} (1-v_{k,t+1})^{\alpha+m_{k,t}-z_{k,t}-1} \\
 & \propto \frac{1}{(\Gamma(z_{k,t}+1))^2 \Gamma(\alpha+m_{k,t}-z_{k,t}) \Gamma(m_{k,t}-z_{k,t}+1)} \left(\frac{v_{k,t} v_{k,t+1}}{(1-v_{k,t})(1-v_{k,t+1})} \right)^{z_{k,t}}.
\end{align*}
Finally, the auxiliary variable for the slice sampler has posterior distribution given by
\begin{equation*}
P(u_{j,t}| \cdots) \propto \left\{
\begin{array}{ll}
\mathbb{I}(u_{j,t} < w_{d_{j,t}})^{\gamma_{j,t}}  	& \mathrm{if} \; \gamma_{j,t}=1 ,\\
\mathbb{I}(u_{j,t} < 1)^{1-\gamma_{j,t}} 			& \mathrm{if} \; \gamma_{j,t}=0.
\end{array} \right.
\end{equation*}

\subsection{Posterior for $\lambda_{j}$}
Posterior distribution for $\lambda_{j}$, for $j=1,\ldots,n^2$. Define $\mu_{k_j}^*,\tau_{k_j}^*$ are the location and scale of $\beta_{j,t}$, respectively, when sparse or non-sparse component is chosen.
\begin{align*}
\notag
P(\lambda_{j}|\cdots) & \propto \lambda_{j}^{-\gamma_{j,t}/2} \exp\left\lbrace -\gamma_{j,t} \frac{(\beta_{j,t} -\mu_{d_{j,t}})^2}{2\lambda_{j}} \right\rbrace \exp\left\lbrace -\gamma_{j,t} \left( \frac{\lambda_{j} \tau_{d_{j,t}}}{2} \right) \right\rbrace \\ \notag
 & \propto \lambda_{j}^{-\gamma_{j,t}/2} \exp\left\lbrace -\frac{1}{2} \left[ \lambda_{j} (\gamma_{j,t} \tau_{d_{j,t}}) + \frac{1}{\lambda_{j}} \left( \gamma_{j,t} (\beta_{j,t}-\mu_{d_{j,t}})^2 \right) \right] \right\rbrace \\
 & \propto GiG(p,a,b)
\end{align*}
where
\begin{equation*}
p= \frac{\gamma_{j,t}}{2}, \qquad a= \gamma_{j,t} \tau_{d_{j,t}}, \qquad b= \gamma_{j,t} (\beta_{j,t}-\mu_{d_{j,t}})^2.
\end{equation*}
The probability density function of a generalised inverse Gaussian, for $p\in \mathbb{R}, a >0, b >0$ and $K_p(\cdot)$ being a modified Bessel function of the second kind, is
\begin{equation*}
GiG(x|p,a,b) = \frac{(a/b)^{p/2}}{2 K_p(\sqrt{ab})} x^{p-1} e^{-(ax+b/x)/2}, \qquad x \in (0,\infty).
\end{equation*}
The vector $\boldsymbol{\lambda} = (\lambda_{1},\ldots,\lambda_{n^2})$ represents the diagonal of the diagonal matrix $\boldsymbol{\Lambda}$ of size $(n^2 \times n^2)$.

\subsection{Posterior of the stick breaking locations $(\mu_{k},\tau_{k})$}
In the non-sparse case, we generate $(\mu_k, \tau_k)$ for $k = 1,\ldots,k^*$ and we have the following full conditional separately for $\mu_k$ and $\tau_k$. The posterior distribution for the stick breaking locations in the non-sparse case, is given by:
\begin{align*}
\notag
P(\mu_{k} | \cdots) & \propto \mathcal{N}(\mu_{k} | c,d) \cdot \prod_{(j,t) : \gamma_{j,t} = 1, d_{j,t} =k} \exp\left\lbrace -\frac{1}{2 \lambda_{j}} (\beta_{j,t} -\mu_{k})^2 \right\rbrace \\ \notag
 & \propto \exp\left\lbrace -\frac{1}{2d} (\mu_{k}^2 -2c\mu_{k}) \right\rbrace \exp\left\lbrace -\frac{1}{2} \sum_{(j,t) : \gamma_{j,t} = 1, d_{j,t}=k} \mu_{k}^2\lambda_{j}^{-1} -2\mu_{k}\lambda_{j}^{-1} \beta_{j,t} \right\rbrace \\ \notag
 & \propto \exp\left\lbrace -\frac{1}{2} \left[ \mu_{k}^2 d^{-1} -2d^{-1}c\mu_{k} + \sum_{(j,t) : \gamma_{j,t} = 1, d_{j,t} = k} \mu_{k}^2 \lambda_{j,t}^{-1} -2\mu_{k} \lambda_{j}^{-1} \beta_{j,t} \right] \right\rbrace \\
 & \propto \mathcal{N}(\overline{c}, \overline{d}),
\end{align*}
where
\begin{equation*}
\overline{d} = \left( d^{-1} + \sum_{(j,t) : \gamma_{j,t} = 1,d_{j,t} = k} \lambda_{j}^{-1} \right)^{-1}, \qquad \overline{c} = \overline{d} \left( d^{-1}c + \sum_{(j,t) : \gamma_{j,t} = 1, d_{j,t} =k} \lambda_{j}^{-1} \beta_{j,t} \right).
\end{equation*}
and for the scale parameter we have the following representation:
\begin{align*}
\notag
P(\tau_{k} | \cdots) & \propto \mathcal{G}a(\tau_{k} | a_1,b_1) \cdot \prod_{(j,t) : \gamma_{j,t}=1, d_{j,t} =k} \frac{\tau_{k}}{2} \exp\left\lbrace -\frac{\tau_{k}}{2} \lambda_{j} \right\rbrace \\ \notag
 & \propto \tau_{k}^{a_1-1} \exp\left\lbrace -\frac{\tau_{k}}{b_1} \right\rbrace \cdot \prod_{(j,t) : \gamma_{j,t}=1,, d_{j,t} =k} \frac{\tau_{k}}{2} \exp\left\lbrace -\frac{\tau_{k}}{2} \lambda_{j} \right\rbrace \\ \notag
 & \propto \tau_{k}^{a_1-1} \exp\left\lbrace -\frac{\tau_{k}}{b_1} \right\rbrace \tau_{k}^{m_1} \exp\left\lbrace -\frac{\tau_{k}}{2} \sum_{(j,t) : \gamma_{j,t}=1, d_{j,t} =k} \lambda_{j} \right\rbrace \\ \notag
 & \propto \tau_{k}^{a_1+m_1-1} \exp\left\lbrace -\tau_{k} \left( \frac{1}{b_1} + \frac{1}{2} \sum_{(j,t) : \gamma_{j,t}=1, d_{j,t} =k} \lambda_{j} \right) \right\rbrace \\
 & \propto \mathcal{G}a(\overline{a}_1, \overline{b}_1),
\end{align*}
with
\begin{equation*}
\overline{a}_1 = a_1 + m_1, \qquad \overline{b}_1 = \frac{2b_1}{2+b_1 \sum_{(j,t) : \gamma_{j,t}=1, d_{j,t} =k} \lambda_{j,t}},
\end{equation*}
where $m_1 = \sum_{(j,t)} \mathbb{I}(\gamma_{j,t}=1) \mathbb{I}(d_{j,t} = k) = \sum_{j=1}^{n^2} \sum_{t=2}^T \gamma_{j,t}\mathbb{I}(d_{j,t} = k)$.

\subsection{Posterior of the allocation variable $(d_{j,t},\gamma_{j,t})$}
Regarding the allocation variables $(d_{j,t},\gamma_{j,t})$, we obtain the following full conditional for the sparse and non-sparse case. The (conditional) prior distribution for the allocation variable $d_{j,t}$, for each $k=1,\dots, k^*$, $j=1,\ldots,n^2$, $t=2,\ldots,T$, is given by:
\begin{align*}
P(d_{j,t} = k | \gamma_{j,t} = 1) & = w_k, \\
P(d_{j,t} = k | \gamma_{j,t} = 0) & = 0, \\
P(d_{j,t} = 0 | \gamma_{j,t} = 1) & = 0, \\
P(d_{j,t} = 0 | \gamma_{j,t} = 0) & = 1,
\end{align*}
while the prior for $\gamma_{j,t}$ is 
\begin{align*}
P(\gamma_{j,t} = 1) & = (1-\pi_t) \\
P(\gamma_{j,t} = 0) & = \pi_t
\end{align*}
Define $\boldsymbol{\gamma}_{-j,t} = (\gamma_{1,t},\ldots,\gamma_{j-1,t},\gamma_{j+1,t},\ldots,\gamma_{n^2,t})'$, $\mathbf{d}_{-j,t} = (d_{1,t},\ldots,d_{j-1,t},d_{j+1,t},\ldots,d_{n^2,t})'$. The joint posterior distribution of the allocation variables $(d_{j,t},\gamma_{j,t})$, for each $k=1,\dots, k^*$, $j=1,\ldots,n^2$, $t=2,\ldots,T$, is obtained as:
\begin{align*}
\notag
P(d_{j,t} = k, \gamma_{j,t} = 1| \cdots) & \propto (1-\pi_t) P(\mathbf{y}_t|d_{j,t} = k, \gamma_{j,t} = 1,\mathbf{d}_{-j,t},\boldsymbol{\gamma}_{-j,t},\dots), \\
P(d_{j,t} = k, \gamma_{j,t} = 0| \cdots) & = 0, \\
P(d_{j,t} = 0, \gamma_{j,t} = 1| \cdots) & = 0, \\
P(d_{j,t} = 0, \gamma_{j,t} = 0| \cdots) & \propto \pi_t P(\mathbf{y}_t|d_{j,t} = 0, \gamma_{j,t} = 0,\mathbf{d}_{-j,t},\boldsymbol{\gamma}_{-j,t},\dots) = \pi_t,
%\mathcal{D}E(\beta_{j,t}| \mu_k,\tau_{k}) \mathbb{I}(u_{j,t} < w_k) \\
% & \propto \frac{\mathcal{D}E(\beta_{j,t}| \mu_{k},\tau_{k})}{\sum_{i\in A_w(u_{j,t})} \mathcal{D}E(\beta_{i,t}| \mu_{i},\tau_{i})}, \quad \forall \, k \in A_w(u_{j,t}),
\end{align*}
where the (conditional) marginal likelihood obtained by integrating out the $\boldsymbol{\beta}_t$, for each $t=2,\ldots,T$, is given by
\begin{align*}
P(\mathbf{y}_t & | \boldsymbol{\gamma}_t, \mathbf{d}_t, \boldsymbol{\lambda}, \boldsymbol{\Sigma}) = \int P(\mathbf{y}_t| \boldsymbol{\beta}_t, \boldsymbol{\Sigma}) P(\boldsymbol{\beta}_t | \boldsymbol{\gamma}_t, \mathbf{d}_t, \boldsymbol{\lambda}) \: \mathrm{d}\boldsymbol{\beta}_t \\
 & = \int (2\pi)^{-n/2} \abs{\boldsymbol{\Sigma}}^{-1/2} \exp\left\lbrace -\frac{1}{2} (\mathbf{y}_t-\mathbf{X}_t \boldsymbol{\beta}_t)' \boldsymbol{\Sigma}^{-1} (\mathbf{y}_t-\mathbf{X}_t \boldsymbol{\beta}_t) \right\rbrace \cdot \left[ \prod_{j:\gamma_{j,t}=0} \delta_{(0)}(\beta_{j,t}) \right] \\
 & \quad \cdot (2\pi)^{-q/2} \abs{\boldsymbol{\Lambda}_{\boldsymbol{\gamma},\mathbf{d},t}}^{-1/2} \exp\left\lbrace -\frac{1}{2} (\boldsymbol{\beta}_{\boldsymbol{\gamma},t} -\boldsymbol{\mu}_{\boldsymbol{\gamma},\mathbf{d},t})' \boldsymbol{\Lambda}_{\boldsymbol{\gamma},t}^{-1} (\boldsymbol{\beta}_{\boldsymbol{\gamma},t} -\boldsymbol{\mu}_{\boldsymbol{\gamma},\mathbf{d},t}) \right\rbrace \: \mathrm{d}\boldsymbol{\beta}_{\mathbf{0},t} \: \mathrm{d}\boldsymbol{\beta}_{\boldsymbol{\gamma},t} \\
 & = (2\pi)^{-n/2} \abs{\boldsymbol{\Sigma}}^{-1/2} \abs{\boldsymbol{\Lambda}_{\boldsymbol{\gamma},\mathbf{d},t}}^{-1/2} \exp\left\lbrace -\frac{1}{2} \left[ \mathbf{y}_t' \boldsymbol{\Sigma}^{-1} \mathbf{y}_t + \boldsymbol{\mu}_{\boldsymbol{\gamma},\mathbf{d},t}' \boldsymbol{\Lambda}_{\boldsymbol{\gamma},\mathbf{d},t}^{-1} \boldsymbol{\mu}_{\boldsymbol{\gamma},\mathbf{d},t} \right] \right\rbrace \\
 & \quad \cdot \int \exp\left\lbrace -\frac{1}{2}\left[ \boldsymbol{\beta}_{\boldsymbol{\gamma},t}' (\mathbf{X}_{\boldsymbol{\gamma},\mathbf{d},t}' \boldsymbol{\Sigma}^{-1} \mathbf{X}_{\boldsymbol{\gamma},\mathbf{d},t} + \boldsymbol{\Lambda}_{\boldsymbol{\gamma},\mathbf{d},t}^{-1}) \boldsymbol{\beta}_{\boldsymbol{\gamma},t} -2(\mathbf{y}_t \boldsymbol{\Sigma}^{-1} \mathbf{X}_{\boldsymbol{\gamma},\mathbf{d},t} + \boldsymbol{\mu}_{\boldsymbol{\gamma},\mathbf{d},t}' \boldsymbol{\Lambda}_{\boldsymbol{\gamma},\mathbf{d},t}^{-1}) \boldsymbol{\beta}_{\boldsymbol{\gamma},t} \right] \right\rbrace \: \mathrm{d}\boldsymbol{\beta}_{\boldsymbol{\gamma},t} \\
 & = (2\pi)^{-\frac{n}{2}} \left( \frac{\abs{\boldsymbol{\Sigma}_{\boldsymbol{\beta}_{\boldsymbol{\gamma},t}}}}{\abs{\boldsymbol{\Sigma}}} \right)^{\frac{1}{2}} \abs{\boldsymbol{\Lambda}_{\boldsymbol{\gamma},\mathbf{d},t}}^{-\frac{1}{2}}  \exp\left\lbrace -\frac{1}{2} \left[ \mathbf{y}_t' \boldsymbol{\Sigma}^{-1} \mathbf{y}_t + \boldsymbol{\mu}_{\boldsymbol{\gamma},\mathbf{d},t}' \boldsymbol{\Lambda}_{\boldsymbol{\gamma},\mathbf{d},t}^{-1} \boldsymbol{\mu}_{\boldsymbol{\gamma},\mathbf{d},t} -\boldsymbol{\mu}_{\boldsymbol{\beta}_{\boldsymbol{\gamma},t}}' \boldsymbol{\Sigma}_{\boldsymbol{\beta}_{\boldsymbol{\gamma},t}}^{-1} \boldsymbol{\mu}_{\boldsymbol{\beta}_{\boldsymbol{\gamma},t}} \right] \right\rbrace,
\end{align*}
where $\mathbf{X}_{\boldsymbol{\gamma},\mathbf{d},t}$ contains the columns of $\mathbf{X}_t$ corresponding to the $j$ such that $\gamma_{j,t}=1$ and $d_{j,t}=\mathbf{d}_j$ (similarly for $\boldsymbol{\Lambda}_{\boldsymbol{\gamma},\mathbf{d},t}$), $\boldsymbol{\mu}_{\boldsymbol{\gamma},\mathbf{d},t}$ is the sub-vector of $(\mu_{d_{1,t}},\ldots,\mu_{d_{n^2,t}})'$ containing the elements $j$ such that $\gamma_{j,t} = 1$, and
\begin{equation*}
\boldsymbol{\Sigma}_{\boldsymbol{\beta}_{\boldsymbol{\gamma},t}} = (\mathbf{X}_{\boldsymbol{\gamma},\mathbf{d},t}' \boldsymbol{\Sigma}^{-1} \mathbf{X}_{\boldsymbol{\gamma},\mathbf{d},t} + \boldsymbol{\Lambda}_{\boldsymbol{\gamma},\mathbf{d},t}^{-1})^{-1}, \qquad
\boldsymbol{\mu}_{\boldsymbol{\beta}_{\boldsymbol{\gamma},t}} = \boldsymbol{\Sigma}_{\boldsymbol{\beta}_{\boldsymbol{\gamma},t}} (\mathbf{X}_{\boldsymbol{\gamma},\mathbf{d},t}' \boldsymbol{\Sigma}^{-1} \mathbf{y}_t + \boldsymbol{\Lambda}_{\boldsymbol{\gamma},\mathbf{d},t}^{-1} \boldsymbol{\mu}_{\boldsymbol{\gamma},\mathbf{d},t}).
\end{equation*}

\subsection{Posterior for $\beta_{j,t}$}
Posterior distribution for $\boldsymbol{\beta}_t$, for all $t=2,\ldots,T$.
In the sparse case we have
\begin{equation*}
P(\beta_{j,t} | \gamma_{j,t}=0, \cdots) = \delta_{(0)}(\beta_{j,t}).
\end{equation*}
In the non-sparse case, denoting $\boldsymbol{\beta}_{\boldsymbol{\gamma},t}$ the sub-vector of $\boldsymbol{\beta}_t$ corresponding to the coefficients such that $\gamma_{j,t}=1$, we have
\begin{align*}
\notag
P(\boldsymbol{\beta}_{\boldsymbol{\gamma},t}&|\cdots)  \propto (2\pi)^{-n^2/2} \abs{\boldsymbol{\Lambda}_{\boldsymbol{\gamma},\mathbf{d},t}}^{-1/2} \exp\left\lbrace -\frac{1}{2} (\boldsymbol{\beta}_{\boldsymbol{\gamma},t}-\boldsymbol{\mu}_{\boldsymbol{\gamma},\mathbf{d},t})' \boldsymbol{\Lambda}_{\boldsymbol{\gamma},\mathbf{d},t}^{-1} (\boldsymbol{\beta}_{\boldsymbol{\gamma},t}-\boldsymbol{\mu}_{\boldsymbol{\gamma},\mathbf{d},t}) \right\rbrace \\ \notag
 & \quad \cdot (2\pi)^{-n^2/2} \abs{\boldsymbol{\Sigma}}^{-1/2} \exp\left\lbrace -\frac{1}{2} (\mathbf{y}_t -\mathbf{X}_{\boldsymbol{\gamma},\mathbf{d},t} \boldsymbol{\beta}_{\boldsymbol{\gamma},t})' \boldsymbol{\Sigma}_{\epsilon}^{-1} (\mathbf{y}_t -\mathbf{X}_{\boldsymbol{\gamma},\mathbf{d},t} \boldsymbol{\beta}_{\boldsymbol{\gamma},t}) \right\rbrace \\ \notag
 & \propto \exp\left\lbrace -\frac{1}{2} \left[ \boldsymbol{\beta}_{\boldsymbol{\gamma},t}' \boldsymbol{\Lambda}_{\boldsymbol{\gamma},\mathbf{d},t}^{-1} \boldsymbol{\beta}_{\boldsymbol{\gamma},t} -2\boldsymbol{\beta}_{\boldsymbol{\gamma},t}' \boldsymbol{\Lambda}_{\boldsymbol{\gamma},\mathbf{d},t}^{-1} \boldsymbol{\mu}_{\boldsymbol{\gamma},\mathbf{d},t} + \boldsymbol{\beta}_{\boldsymbol{\gamma},t}' \mathbf{X}_{\boldsymbol{\gamma},\mathbf{d},t}' \boldsymbol{\Sigma}^{-1} \mathbf{X}_{\boldsymbol{\gamma},\mathbf{d},t} \boldsymbol{\beta}_{\boldsymbol{\gamma},t} -2\boldsymbol{\beta}_{\boldsymbol{\gamma},t}' \mathbf{X}_{\boldsymbol{\gamma},\mathbf{d},t}' \boldsymbol{\Sigma}^{-1} \mathbf{y}_t \right] \right\rbrace \\
 & \propto \mathcal{N}(\boldsymbol{\mu}_{\boldsymbol{\beta}_{\boldsymbol{\gamma},t}},\boldsymbol{\Sigma}_{\boldsymbol{\beta}_{\boldsymbol{\gamma},t}}),
\end{align*}
where
\begin{equation*}
\boldsymbol{\Sigma}_{\boldsymbol{\beta}_{\boldsymbol{\gamma},t}} = (\mathbf{X}_{\boldsymbol{\gamma},\mathbf{d},t}' \boldsymbol{\Sigma}^{-1} \mathbf{X}_{\boldsymbol{\gamma},\mathbf{d},t} + \boldsymbol{\Lambda}_{\boldsymbol{\gamma},\mathbf{d},t}^{-1})^{-1}, \qquad
\boldsymbol{\mu}_{\boldsymbol{\beta}_{\boldsymbol{\gamma},t}} = \boldsymbol{\Sigma}_{\boldsymbol{\beta}_{\boldsymbol{\gamma},t}} (\mathbf{X}_{\boldsymbol{\gamma},\mathbf{d},t}' \boldsymbol{\Sigma}^{-1} \mathbf{y}_t + \boldsymbol{\Lambda}_{\boldsymbol{\gamma},\mathbf{d},t}^{-1} \boldsymbol{\mu}_{\boldsymbol{\gamma},\mathbf{d},t}).
\end{equation*}

\subsection{Posterior for covariance matrix $\Sigma$}
Posterior distribution for the covariance matrix $\boldsymbol{\Sigma}$.
\begin{align*}
\notag
P(\boldsymbol{\Sigma} | \cdots) & \propto \mathcal{IW}(\nu,\boldsymbol{\Psi}) \cdot \prod_{t=2}^T \abs{\boldsymbol{\Sigma}}^{-1/2} \exp\left\lbrace -\frac{1}{2} (\mathbf{y}_t - \mathbf{X}_t \boldsymbol{\beta}_t)' \boldsymbol{\Sigma}^{-1} (\mathbf{y}_t - \mathbf{X}_t \boldsymbol{\beta}_t) \right\rbrace \\ \notag
 & \propto \abs{\boldsymbol{\Sigma}}^{-\frac{\nu+n+1}{2}} \exp\left\lbrace -\frac{1}{2} \operatorname{tr}\left( \boldsymbol{\Psi} \boldsymbol{\Sigma}^{-1} \right) \right\rbrace \abs{\boldsymbol{\Sigma}}^{-\frac{T-1}{2}} \exp\left\lbrace -\frac{1}{2} \sum_{t=2}^T \operatorname{tr}\left( (\mathbf{y}_t - \mathbf{X}_t \boldsymbol{\beta}_t)' \boldsymbol{\Sigma}^{-1} (\mathbf{y}_t - \mathbf{X}_t \boldsymbol{\beta}_t) \right) \right\rbrace \\
 & \propto \mathcal{IW}(\overline{\nu}, \overline{\boldsymbol{\Psi}}),
\end{align*}
where
\begin{equation*}
\overline{\nu} = \nu+T-1, \qquad \overline{\boldsymbol{\Psi}} = \boldsymbol{\Psi} + \sum_{t=2}^T (\mathbf{y}_t - \mathbf{X}_t \boldsymbol{\beta}_t)(\mathbf{y}_t - \mathbf{X}_t \boldsymbol{\beta}_t)'.
\end{equation*}

\subsection{Posterior for mixing probability $\pi_t$}
Posterior distribution for the mixing probability $\pi_t$, for all $t=2,\ldots,T$.
\begin{align*}
\notag
P(\pi_t | \cdots) & \propto \mathcal{B}e(1,\eta) \cdot \prod_{j=1}^{n^2} \pi_t^{\mathbb{I}(\gamma_{j,t}=0)} (1-\pi_t)^{\mathbb{I}(\gamma_{j,t}=1)} \\ \notag
 & \propto (1-\pi_t)^{\eta-1} \: \pi_t^{n^2-\sum_{j=1}^{n^2} \mathbb{I}(\gamma_{j,t}=1)} \: (1-\pi_t)^{\sum_{j=1}^{n^2} \mathbb{I}(\gamma_{j,t}=1)} \\
 & \propto \mathcal{B}e(\overline{\phi}, \overline{\eta}),
\end{align*}
where
\begin{equation*}
\overline{\phi} = 1+n^2 -\sum_{j=1}^{n^2} \mathbb{I}(\gamma_{j,t}=1), \qquad \overline{\eta} = \eta +\sum_{j=1}^{n^2} \mathbb{I}(\gamma_{j,t}=1).
\end{equation*}

\end{document}